# Implications of coordination chemistry to cationic interactions in honeycomb layered nickel tellurates

Kohei Tada[a], Titus Masese[a,b] & Godwill Mbiti Kanyolo[c]

[a] Research Institute of Electrochemical Energy, National Institute of Advanced Industrial Science and Technology (AIST), 1–8–31 Midorigaoka, Ikeda, Osaka 563–8577, JAPAN
[b] AIST–Kyoto University Chemical Energy Materials Open Innovation Laboratory (ChEM–OIL), Sakyo–ku, Kyoto 606–8501, JAPAN
[c] Department of Engineering Science, The University of Electro-Communications, 1–5–1 Chofugaoka, Chofu, Tokyo 182–8585, JAPAN

## Highlights

- Density-functional theory calculations for $A_2Ni_2TeO_6$ ($A$ = Rb, Cs, H, Cu, Ag and Au).
- Large alkali atoms (Na, Rb and Cs) form a prismatic coordination while Li forms octahedral coordination with oxygen atoms at large and small interlayer distances respectively.
- Coinage-metal atoms (Cu, Ag and Au) form a linear coordination with oxygen atoms at intermediate interlayer distances.
- Hydrogen atoms form linear coordinations with oxygen atoms forming hydroxyl radical groups.
- A linear coordination, coupled with the presence of metallophilic bonds, offers a possibility for novel cationic structures (such as bilayers) in honeycomb layered frameworks.

## Abstract

Honeycomb layered tellurates represent a burgeoning class of multi-functional materials with fascinating crystal-structural versatility and a rich composition space. Despite their multifold capabilities, their compositional diversity remains underexplored due to complexities in experimental design and syntheses. Thus, in a bid to expand this frontier and derive relevant insights into allowed metastable compositions, we employ a density functional theory (DFT) approach to predict *in silico* the crystal structures of new honeycomb layered tellurates embodied by the composition, $A_2Ni_2TeO_6$ ($A$ = alkali, hydrogen or coinage-metal cations). Here, alkali-metal atoms with vastly larger radii than K (Rb and Cs) are found to engender a prismatic coordination with the oxygen atoms from the honeycomb slabs whilst coinage-metal atoms (Ag, Au and Cu) display a propensity for linear coordination. Further, $H_2Ni_2TeO_6$ is found to also render a linear coordination wherein the hydrogen atom preferentially establishes a stronger coordination with one of the oxygen atoms to form hydroxyl groups. All *A* cations in the studied $A_2Ni_2TeO_6$ compositions form a honeycomb lattice. Conclusions on the possibility of a monolayer-bilayer phase transition in coinage metal atom tellurates can be drawn by considering the implications of conformal symmetry of the cation honeycomb lattice and metallophilicity. This work not only propounds new honeycomb layered tellurate compositions but also provides novel insight into the rational design of multifunctional materials for applications ranging from energy storage, catalysis and optics to analogue condensed matter systems of gravity.

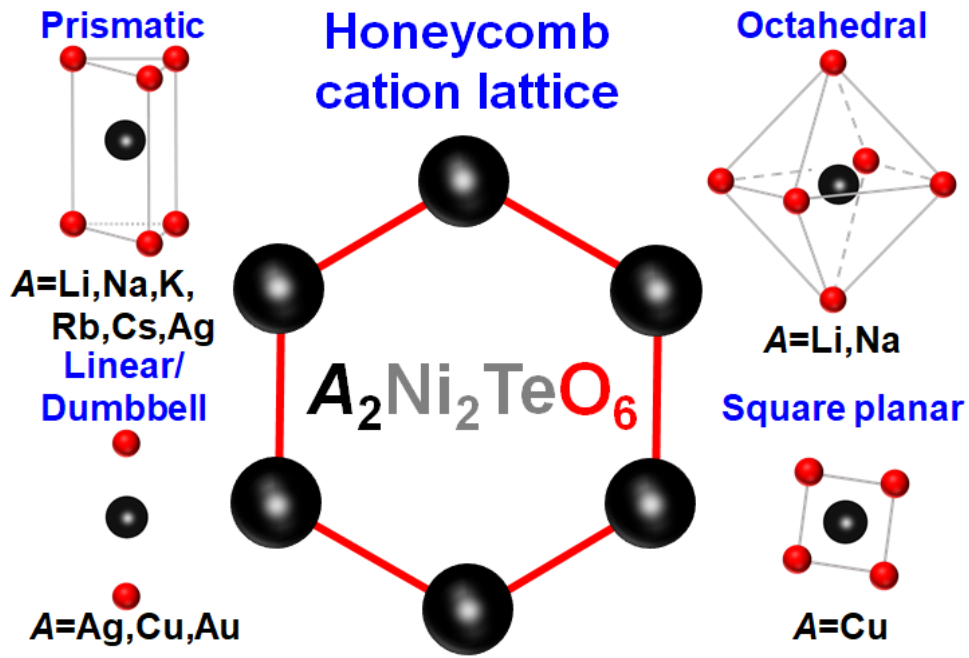

**Rendition:** The honeycomb lattice of cations in $A_2Ni_2TeO_6$ and their respective coordinations with oxygen atoms as found by Density Functional Theory (DFT) calculations

# **INTRODUCTION**

Honeycomb layered oxides embodying monovalent atoms such as Li, Na, K, Cu, or Ag, encapsulated between layered hexagonal frameworks of transition metal- or heavy metal oxides, exemplify a vanguard class of nanomaterials endowed with unconventional magnetic interactions, agile ionic mobility, exotic phase transitions and exquisite electrochemical functionalities that betoken monumental advancements in science and technology.[1] These fascinating capabilities have galvanised inquests into honeycomb layered oxides, particularly those entailing pnictogen and chalcogen atom frameworks, for they not only harbour a rich compositional canvas for developing high-voltage materials with high ionic conductivities but also represent an ideal pedagogical platform for investigating piquant electromagnetic phenomena and analogue gravity systems.[1-45]

At the heart of these exotic phenomena is coordination chemistry and crystalline symmetry, which harbour a great deal of structural, thermodynamic and quantum information relevant to unearthing the various mechanisms within honeycomb layered oxides.[1] For instance, it is known that the coordination of cations with the oxygen in the slabs is modified by cationic vacancy annihilation and creation during the intercalation and de-intercalation processes respectively, occurring during cell cycling when these materials are used as cathodes.[1] Meanwhile, the cations are expected to form a honeycomb lattice, spanned by rhombic unit cells which can be mapped onto other unit cells by modular/conformal symmetry, implying some underlying conformal field theory describes the dynamics of such cations.[78,79] Consequently, an idealised model of cationic diffusion has suggested that the cationic vacancies can be treated as the genus of an emergent two-dimensional orientable manifold without boundary, where the dynamics of the cations is governed by Chern-Simons theory in one time and two spatial (*1, 2*) dimensions, whereas the dynamics of the cationic vacancies is a gravitational theory in *1, 3* dimensions.[1,43-45] Recent developments in theoretical physics has linked certain classes of conformal field theories (CFTs) in *1, d* dimensions to quantum geometry theories in *1, d+1* dimensions, suggesting that honeycomb layered oxides are exemplars of the so-called gravity/CFT duality.[78,79] This offers a tool to analyse the link between coordination chemistry and cationic vacancies by finding stable geometric configuration solutions in the gravity dual, whereby various atoms in the material can rearrange themselves in what we refer to as *Jenga mechanism*.[1]

Honeycomb layered oxides comprising pnictogen and/or chalcogen frameworks adopt an assortment of chemical compositions such as $A_4M$DO$_6$, $A_3M_2$DO$_6$ or $A_2M_2$DO$_6$, *inter alia*; where *M* represents transition metal species such as Cu, Co, Zn, Ni, *etc.* or *s*-block metals such as Mg; *A* denotes alkali- or coinage-metal species *e.g.*, Li, Na, K, Cu, Ag, *etc*., and *D* depicts a chalcogen or pnictogen metal species such as Te, Sb, Bi, *etc*. [1,2,6,8,10-12,15,18-21,29-31] These compounds present a rich compositional flexibility that allows various combinations and permutations of the *M* and *D* species—a recipe for unique magnetic and voltage properties that further expand their range of applications. Despite this chemical diversity, exploration into chalcogen *D* species such as Te, remains limited, considering the great deal of scientific endeavour dispensed into honeycomb layered oxides entailing pnictogens such as Bi and Sb. [2,8,11,12-14,22,24,27,30]

Notably, honeycomb layered tellurates adopting the $A_2M_2$TeO$_6$ (*A* = Li, Na, K) compositions have garnered great utility as energy storage materials for they exhibit superb ionic properties such as elevated conductivities as well as high voltages that undoubtedly supersede those of most antimonates and bismuthates.[1] In fact, tellurates used alongside Ni in $A_2$Ni$_2$TeO$_6$ honeycomb layered oxides have been reported to exhibit the highest ionic conductivities and voltages (over 4 V) amongst the honeycomb layered oxides reported to date.[1,17,19-21,23,26,41,42] Therefore, to further expand this compositional space and enhance the functionalities of nickel-based honeycomb layered tellurates, it is imperative to gain insights into related layered structures that can accommodate other alkali atoms — (*e.g.,* Rb and Cs) or coinage metal atoms (such as Ag and Cu) —a pursuit yet to be undertaken for these tellurates.

In the exemplar case of Ag, preliminary experimental results show a pair of triangular lattices of cations stacked on each other, whereby the honeycomb lattice appears to bifurcate into two triangular sub-lattices. [79] This serendipitous result has been theoretically attributed to the breaking of the aforementioned conformal symmetry, whereby the requisite conditions for the monolayer-bilayer phase transition are considered to be a dumbbell/linear coordination coupled with the presence of a finite argentophilic attraction which stabilises the bilayer. [79] At the critical point where conformal symmetry of the cations is restored, the gravity dual describing the cationic vacancies is predicted to be a hyperbolic geometry known as Anti-de Sitter space. [79] However, the experimental discovery, design, and development of novel materials is heavily encumbered by challenges in synthesis and inexplicable influences of external/environmental factors. For this reason, theoretical and computational techniques have become an indispensable part of material science research for their efficacy in not only predicting new materials and their emergent properties under controlled conditions that circumvent the limitations of experimental routes but also effectively analysing metastable configurations which sit at theoretical critical points where phase transitions occur.[83] In particular, density functional theoretical (DFT) methods have garnered traction in manifold applications due to their predictive power rooted in the quantum-mechanical description of electron and atom interactions that not only facilitates the exploration of mesoscopic and microscopic properties but also the prediction of new material compositions alongside their properties down to the atomic scale.[46] As such, DFT has become a uniquely potent tool for predicting new crystal structures with high fidelity besides reproducing material ground state properties with reasonable scaling in respect to system size.

In this study, we not only augment the compositional space of nickel-based honeycomb layered tellurates by employing DFT calculations to predict original compositions of $A_2Ni_2TeO_6$ ($A$ = H, Rb, Cs, Cu, Ag, Au) yet under-reported in experimental and theoretical literature, but also consider the implications of coordination chemistry to cationic interactions beyond alkali-metal-based honeycomb layered nickel tellurates. The computations elucidate the stability and structural configurations of these new compositions comprising vastly large alkali atoms (namely, Rb and Cs), coinage-metal atoms (namely, Ag, Au, and Cu) and hydrogen. The optimised models of the $A_2Ni_2TeO_6$ compositions derived from the computations reveal distinctive prismatic coordinations between large alkali ($A$ = Rb, Cs) and oxygen atoms, whilst the coinage-metal ($A$ = Ag, Au, Cu) and the hydrogen atoms ($A$ = H) exhibit an inclination towards dumbbell-like linear coordinations with two apical oxygen atoms from the transition metal slabs. Further, the predicted crystal model of the $H_2Ni_2TeO_6$ shows that each hydrogen atom preferentially establishes a stronger coordination with one of the oxygen atoms compared to those encompassing coinage-metal atoms, thus forming a relatively more stable hydroxyl group. By incorporating the known chemistry of 1) metallophilicity and possible sub-valencies of coinage-metal atoms[76,77,80-84]; our computational results imply that the found configurations of the coinage metal oxide tellurates satisfying the additional conditions: 2) a honeycomb lattice of cations; and 3) a linear/dumbbell coordination of cations with oxygen atoms; harbour the possibility for the observation of a monolayer-bilayer phase transition. Thus, the present findings not only envisage the vast compositional space of honeycomb layered tellurates but also paves the way for the design and synthesis of these new multifunctional materials that promise to expand the functionalities and applications of honeycomb layered oxides.

# 1. COMPUTATIONAL METHODOLOGY

The charge density and total energies were optimised to be self-consistent (with a threshold of $10^{-7}$ eV) using the Kohn-Sham formalism. [47] The Perdew-Burke-Ernzerhof expression (GGA-PBE), [48] which is one of the generalised gradient approximations, was adopted as the exchange-correlation functional. Further, a dispersion force correction of the DFT-D3 with Becke-Jonson damping method was adopted to accurately calculate the honeycomb oxide interlayer distance. [49,50] The inner core region was assessed using the projector-augmented-wavefunction method. [51] The number of valency electrons was set as follows: H ($1s^1$), Li($2s^1$), Na ($[3s^1$), K ($3p^64s^1$), Rb ($4p^65s^1$), Cs ($5s^25p^66s^1$), Cu ($3d^{10}4s^1$), Ag ($4d^{10}5s^1$), Au ($5d^{10}6s^1$), Ni ($3d^84s^2$), Te($5s^25p^4$) and O ($2s^22p^4$). An on-site Coulomb correction (DFT+$U$ method) was adopted, [52] with $U = 4$ eV added to the $3d$ orbital of Ni. The electronic spins on the Ni sites coupled in an anti-parallel manner can cause spin contamination error. Although the effect of the error in the total energy was investigated by the approximated spin projection scheme for DFT, [53-55] the attained results do not affect the discussion presented herein. The DFT calculations were performed by Vienna *Ab initio* Simulation Package (VASP) programme. [56-59]

Structural optimisation was performed for all atoms, lattice constants, and cell dimensions. The crystal structures were relaxed until the threshold was reduced below $10^{-5}$ eV or $10^{-2}$ eV / Å. It is prudent to mention that the honeycomb $[Ni_2TeO_6]^{2-}$ slab was maintained in all the initial structures used, regardless of the various adjustments made to the atomic positions and lattice constants of the initial $K_2Ni_2TeO_6$ structures. Moreover, no symmetry restrictions were imposed during the structural optimisation. The energy cut-off was set at 500 eV (wave function) and 2400 eV (charge). $k$-sampling was performed with a 5×5×3 of Γ-centred mesh. **Figure 1** shows the optimised crystal structure of $K_2Ni_2TeO_6$, which was subjected to structural relaxation in order to assess its stability in accommodating other alkali atoms (*viz*., H, Li, Na, Rb, Cs) and coinage-metal atoms (Cu, Ag and Au). The optimised lattice parameters of $K_2Ni_2TeO_6$ are in accord with those previously reported. [23,73] The optimised lattice parameters for $K_2Ni_2TeO_6$ along with the optimised structures ($A_2Ni_2TeO_6$ ($A$ = H, Li, Na, Rb, Cs, Ag, Au and Cu)) have been furnished in **Tables S1**, **S2** and **S3**. Moreover, the interatomic distances of the cations which are situated in a honeycomb lattice are provided in **Table S4**. The most stable electronic configurations for $A_2Ni_2TeO_6$ ($A$ = Li, Na, Rb, Cs, Ag, Au and Cu) were found to be antiferromagnetic, whilst a ferromagnetic state was found for $H_2Ni_2TeO_6$ (see **Tables S5** and **S6**). Although triclinic structures could be obtained, for the sake of a systematic discussion, the discussion hereafter is based on the stable optimised configurations of the honeycomb structures. Renditions of the crystal structures and electron density distributions were performed using VESTA crystallographic software. [60]

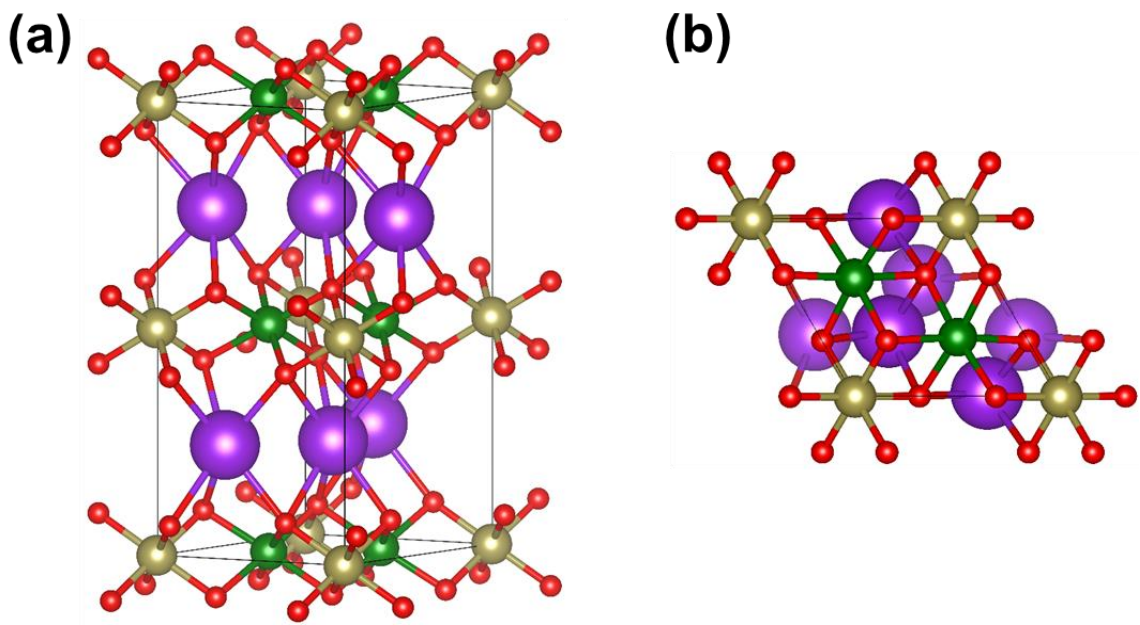

**Figure 1. Optimised crystal structural framework of $K_2Ni_2TeO_6$.** Optimised crystal structural framework of $K_2Ni_2TeO_6$ when viewed along **(a)** the [110] zone axis and **(b)** [001] zone axis. Potassium atoms (in purple) are coordinated with six oxygen atoms (in red) to form a prismatic coordination. The potassium atoms are sandwiched between layers or slabs entailing Ni atoms (dark green) arranged in a honeycomb configuration around Te atoms (in ochre) via the oxygen atoms to form $NiO_6$ and $TeO_6$ octahedra. Note that the crystal structural framework is slightly deviated from the [110] zone axis, in order to explicitly visualise all the atom coordination within the unit cell.

## 2. RESULTS AND DISCUSSION

### 2.1. Crystal structures of $Na_2Ni_2TeO_6$, $Rb_2Ni_2TeO_6$, $Cs_2Ni_2TeO_6$ and $Li_2Ni_2TeO_6$

$K_2Ni_2TeO_6$, a plausible precursor material in the topochemical ion-exchange synthesis of honeycomb layered oxides such as $Na_2Ni_2TeO_6$, was selected for DFT analyses as it displays the widest interlayer distance amongst the nickel-based tellurates reported so far. As such, its lattice structure was assessed to ascertain its disposition to accommodate $Na^+$. An optimised crystal structure of $Na_2Ni_2TeO_6$ was subsequently obtained, as illustrated in **Figures 2a** and **2b**. The derived lattice parameters furnished in **Table S2** (**Supporting Information** section) exhibit congruence with the experimental values—affirming the exactitude of the optimised structure. In the crystal model, each Na atom (in yellow) is coordinated with six oxygen atoms (shown in red) from the adjacent transition metal slabs/layers forming a single $NaO_6$ prismatic polyhedron. On the other hand, each transition metal slab comprises six Ni atoms (shown in grey) surrounding each Te atom (shown in mustard yellow) in a honeycomb fashion (**Figure 2b**). The Na atoms occupy crystallographic sites below the apical oxygen atoms (connecting two Ni atoms and one Te atom in the honeycomb slab), forming a prismatic configuration. This positioning of Na atoms is attributed to the strong electrostatic repulsion imparted by the $Ni^{2+}-Ni^{2+}$ and $Te^{6+}-Te^{6+}$ residing in the adjacent metal slabs due to shorter interslab distance, which prevents them from occupying crystallographic sites where the electrostatic effect of Te and Ni atoms is immense. Other stable configurations, wherein Na is situated directly below or above Ni atoms, were also observed (shown in **Figures 2c** and **2d**). **Figures 2e** and **2f** show another stable configuration of $Na_2Ni_2TeO_6$ with Na atoms coordinated octahedrally with six oxygen atoms. The total energy differences between the structures shown in **Figure 2a** and **Figure 2c** was found to be –0.24 eV, implying the structure shown in **Figure 2a** to be more stable. However, the total energy difference between the structure in **Figure 2a** and that shown in **Figure 2e** was found to be subtle (0.07 eV), indicating the possibility to attain a stable $Na_2Ni_2TeO_6$ structure displaying an octahedral arrangement of Na with oxygen atoms. Given the smaller Shannon-Prewitt ionic radius of $Na^+$ compared to that of $K^+$,[61] the $Na_2Ni_2TeO_6$ crystal renders a smaller interslab/interlayer distance than that of $K_2Ni_2TeO_6$. This is further corroborated by the *c*-axis parameter of $Na_2Ni_2TeO_6$ (**Table S2**), which is lower than that of $K_2Ni_2TeO_6$ (**Table S1**).

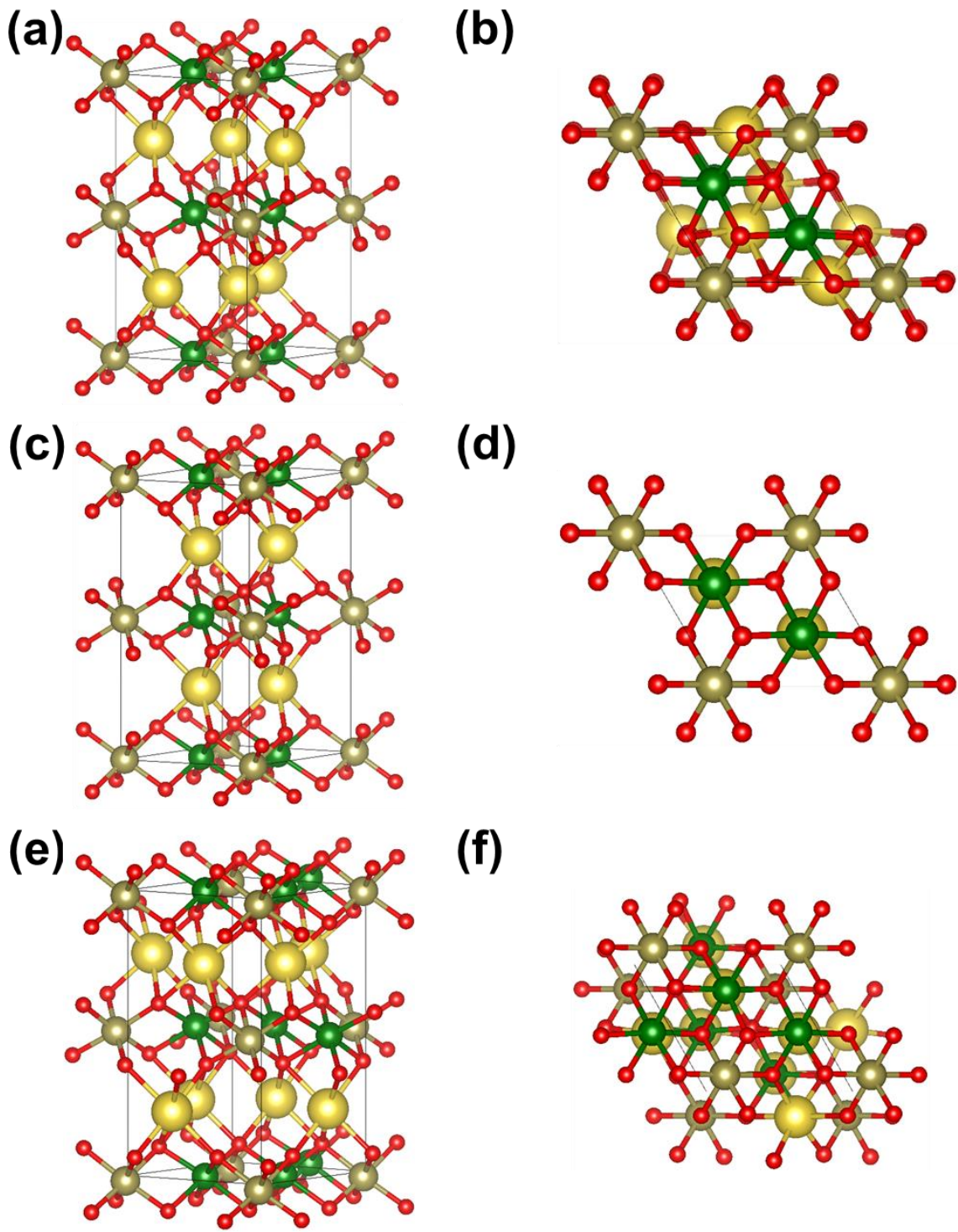

**Figure 2. Optimised crystal structural frameworks of $Na_2Ni_2TeO_6$.** Optimised crystal structural framework of $Na_2Ni_2TeO_6$ (with Na coordinated with oxygen in a prismatic coordination and lying below oxygen atoms (in red)) when viewed along **(a)** [110] zone axis and **(b)** [001] zone axis. Sodium atoms are shown in yellow. **(c)** Optimised crystal structural framework of $Na_2Ni_2TeO_6$ (with Na coordinated with oxygen in a prismatic coordination but lying directly below Ni atoms (in dark green)) when viewed along [110] zone axis and **(d)** when viewed along [001] zone axis. **(e)** Optimised crystal structural framework of $Na_2Ni_2TeO_6$ (with Na coordinated with oxygen in an octahedral coordination) when viewed along [110] zone axis and **(f)** when viewed along [001] zone axis. Ni atoms (dark green) are arranged in a honeycomb configuration around Te atoms (in ochre) via the oxygen atoms to form $NiO_6$ and $TeO_6$ octahedra. Note that the crystal structural framework is slightly deviated from the [110] zone axis, in order to explicitly visualise all the atom coordination within the unit cell. The electron density distribution maps of $Na_2Ni_2TeO_6$ along with the projected density of states (PDOS) plots have been availed in **Figures S1** and **S2.**

The stability and the structural configuration of the $K_2Ni_2TeO_6$ crystal lattice with the larger Rb and Cs atoms was further investigated by obtaining optimised crystal structures of $Rb_2Ni_2TeO_6$ (**Figures 3a** and **3c**) and $Cs_2Ni_2TeO_6$ (**Figures 4a** and **4c**). The pertinent lattice parameters have been furnished in **Table S2**. In both structures, Rb and Cs atoms are coordinated with six oxygen atoms from the adjacent transition metal slabs to form prismatic $RbO_6$ and $CsO_6$ polyhedra (**Figures 2c** and **2e**), in a similar manner as the $Na_2Ni_2TeO_6$ (see **Figure 2a**) and $K_2Ni_2TeO_6$ (**Figure 1a**). Moreover, two stable configurations were attained whereby Rb and Cs atoms occupy crystallographic sites that have Ni atoms (**Figures 3a** and **4a**) or O atoms (**Figures 3c** and **4c**) lying directly above or below. The total energy differences between the obtained structures shown in

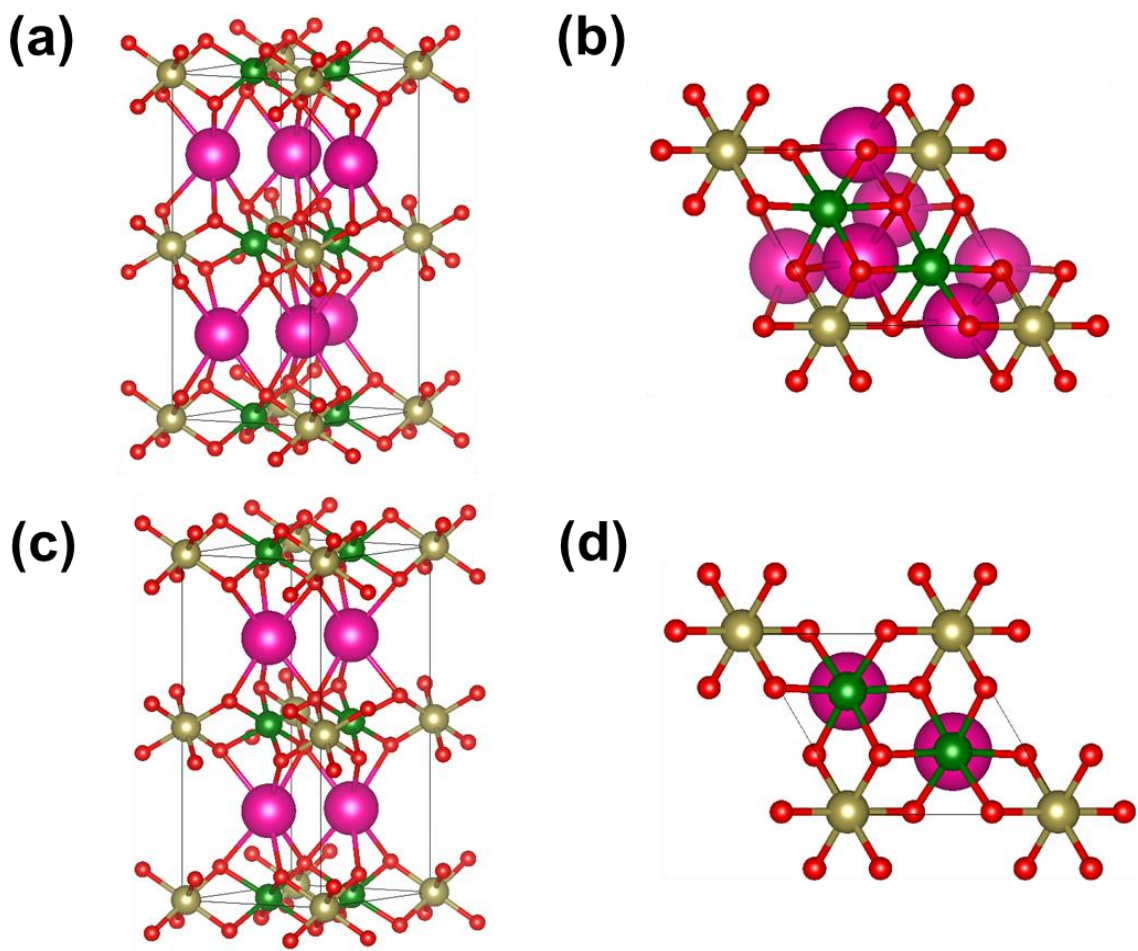

**Figure 3. Optimised crystal structural frameworks of $Rb_2Ni_2TeO_6$.** Optimised crystal structural framework of $Rb_2Ni_2TeO_6$ (with Rb lying below oxygen atoms (in red)) when viewed along **(a)** [110] zone axis and **(b)** [001] zone axis. Rubidium atoms are shown in pink. **(c)** Optimised crystal structural framework of $Rb_2Ni_2TeO_6$ (with Rb situated directly below Ni atoms (in dark green)) when viewed along [110] zone axis and **(d)** when viewed along [001] zone axis. Ni atoms are arranged in a honeycomb configuration around Te atoms (in ochre) via the oxygen atoms to form $NiO_6$ and $TeO_6$ octahedra. Note that the crystal structural framework is slightly deviated from the [110] zone axis, in order to explicitly visualise all the atom coordination within the unit cell. The electron density distribution maps of $Rb_2Ni_2TeO_6$ along with the projected density of states (PDOS) plots have been availed in **Figures S3** and **S4**.

**Figures 3a** and **3c** for $Rb_2Ni_2TeO_6$, and **Figures 4a** and **4c** for $Cs_2Ni_2TeO_6$ were 0.17 eV, and –0.09 eV, respectively. When coordinated with six oxygen atoms, Rb and Cs manifest larger Shannon-Prewitt ionic radii (*i.e.*, 1.52 Å for Rb and 1.67 Å for Cs) than the K atoms (1.38 Å), [61] which endow the $Rb_2Ni_2TeO_6$ and $Cs_2Ni_2TeO_6$ structures with larger interslab/interlayer distances than $K_2Ni_2TeO_6$ (see *c*-axis lattice parameters shown in **Table S2**). Similar to the Na occupancy previously observed (**Figure 2b**), configurations were obtained where the Rb and Cs atoms are positioned directly above and below the Ni atoms in a honeycomb configuration when observed along the *ab* plane (**Figure 3b** and **Figure 4b**). This positioning can be ascribed to the larger interslab distances of $Rb_2Ni_2TeO_6$ and $Cs_2Ni_2TeO_6$ which offset the electrostatic repulsion forces imparted on the alkali atoms by $Ni^{2+}-Ni^{2+}$.

It is worth mentioning that the $Rb_2Ni_2TeO_6$ and $Cs_2Ni_2TeO_6$ are not ideally suited for energy storage applications, despite their projected high voltages, due to the high molar mass of Rb and Cs that leads to significantly low capacities. Even so, their wide interlayer distances are poised to engender unique two-dimensional (2D) magnetic interactions, [1] making these compositions auspicious platforms for related investigations. Additionally, if experimentally synthesised, the most stable configurations of $Rb_2Ni_2TeO_6$ and $Cs_2Ni_2TeO_6$ are postulated to be antiferromagnetic (**Tables S2, S5, and S6**), rendering them promising candidates for exploring emergent 2D antiferromagnetic interactions such as those manifested in the Heisenberg-Kitaev model. [1, 28]

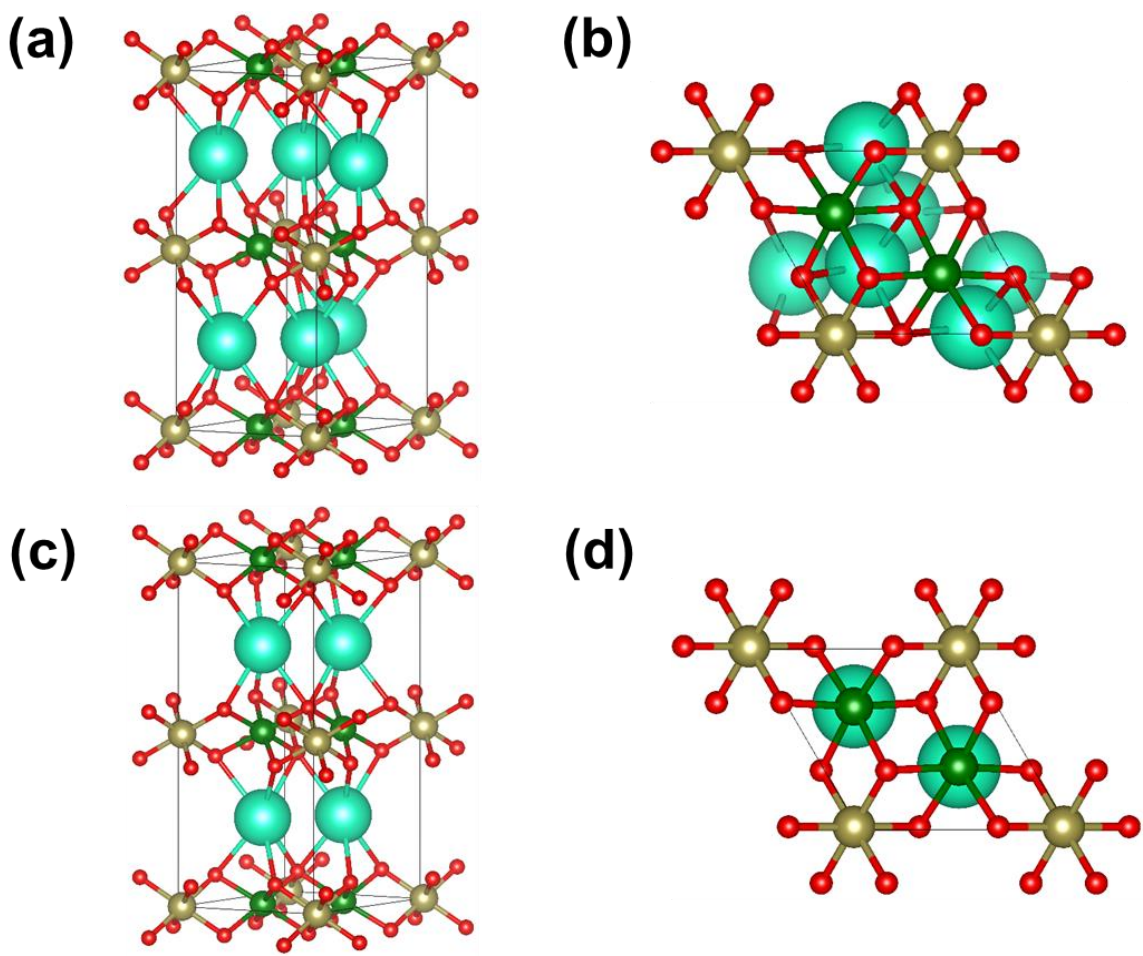

**Figure 4. Optimised crystal structural frameworks of $Cs_2Ni_2TeO_6$.** Optimised crystal structural framework of $Cs_2Ni_2TeO_6$ (with Rb lying below oxygen atoms (in red)) when viewed along **(a)** [110] zone axis and **(b)** [001] zone axis. Caesium atoms are shown in cyan. **(c)** Optimised crystal structural framework of $Cs_2Ni_2TeO_6$ (with Cs situated directly below Ni atoms (in dark green)) when viewed along [110] zone axis and **(d)** when viewed along [001] zone axis. Ni atoms are arranged in a honeycomb configuration around Te atoms (in ochre) via the oxygen atoms to form $NiO_6$ and $TeO_6$ octahedra. Note that the crystal structural framework is slightly deviated from the [110] zone axis, in order to explicitly visualise all the atom coordination within the unit cell. The electron density distribution maps of $Cs_2Ni_2TeO_6$ along with the projected density of states (PDOS) plots have been availed in **Figures S5** and **S6**.

Following the assessments on the crystallographic occupation of the large Rb and Cs atoms in the $K_2Ni_2TeO_6$ crystal lattice, the crystal structure was investigated with Li atoms, which definitively display smaller ionic radii than the K and Na atoms (see the *c*-axis lattice parameter of $Li_2Ni_2TeO_6$ in **Table S2**). The optimised crystal structures of $Li_2Ni_2TeO_6$, viewed along various zone axes, are illustrated in **Figure 5**. In the same vein as the previous alkali atoms, Li atoms are coordinated with six oxygen atoms from the adjacent metal slabs to form $LiO_6$ with Li residing in crystallographic sites where Ni atoms (**Figures 5a** and **5b**) or oxygen atoms (**Figures 5c** and **5d**) lie directly below and above. Furthermore, an optimised crystal structure where Li atoms are coordinated octahedrally with oxygen atoms was also obtained (**Figures 5e** and **5f**). The total energy difference between the structures shown in **Figure 5a** and **Figure 5c** is –0.45 eV, which means that the structure shown in **Figure 5a** is the more stable. However, the total energy difference between the structures shown in **Figure 5a** and **Figure 5e** yields a value of 0.77 eV, indicating that the structure shown in **Figure 5e** (where Li atoms are coordinated octahedrally to oxygen atoms) to be the most stable. Unlike K, Rb and Cs atoms which manifest prismatic coordination of alkali atoms as stable configurations, the Li atoms engender an octahedral coordination with the oxygen atoms (as also noted in a stable configuration of $Na_2Ni_2TeO_6$). Due to the small ionic radius of Li (0.76 Å), [61] the $LiO_6$ octahedra (**Figure 5e**) exhibits a smaller interslab/interslab distance with the most stable bonds among the alkali atoms investigated herein. The reason underlying the stable octahedral coordination of $LiO_6$ in $Li_2Ni_2TeO_6$ can be ascribed as follows. When Li is coordinated with oxygen atoms, their 2*s* and 2*p* orbitals take part in the coordination chemistry. Given that the radial distribution of the orbitals tends to hybridise when in close proximity, the orthogonality of the $p_x$, $p_y$ and $p_z$ orbitals will be reflected by the predominant octahedral coordination of $LiO_6$. To accommodate the $LiO_6$ octahedra in the honeycomb structure, the adjacent transition metal slabs shift across the *ab* plane placing the apical oxygen atoms (connecting two Ni atoms and one Te atom in the honeycomb slab) directly above and below the Li atoms crystallographic sites (**Figures 5e** and **5f**).

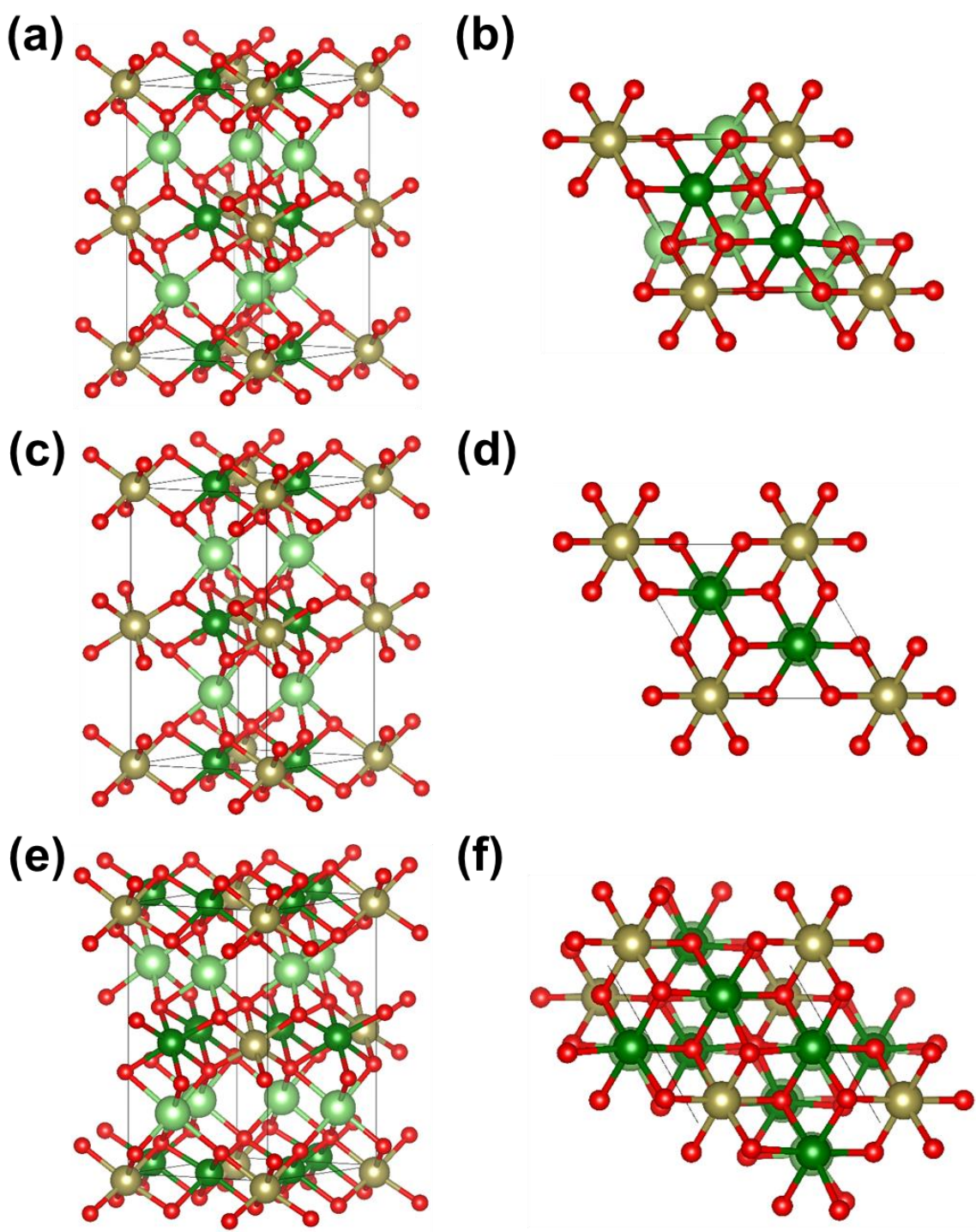

**Figure 5. Optimised crystal structural frameworks of Li$_2$Ni$_2$TeO$_6$.** Optimised crystal structural framework of Li$_2$Ni$_2$TeO$_6$ (with Li coordinated with oxygen in a prismatic coordination and lying below oxygen atoms (in red)) when viewed along **(a)** [110] zone axis and **(b)** [001] zone axis. Lithium atoms are shown in light green. **(c)** Optimised crystal structural framework of Li$_2$Ni$_2$TeO$_6$ (with Li coordinated with oxygen in a prismatic coordination but lying directly below Ni atoms (in dark green)) when viewed

along [110] zone axis and **(d)** when viewed along [001] zone axis. **(e)** Optimised crystal structural framework of $Li_2Ni_2TeO_6$ (with Li coordinated with oxygen in an octahedral coordination) when viewed along [110] zone axis and **(f)** when viewed along [001] zone axis. Ni atoms are arranged in a honeycomb configuration around Te atoms (in ochre) via the oxygen atoms to form $NiO_6$ and $TeO_6$ octahedra. Note that the crystal structural framework is slightly deviated from the [110] zone axis, in order to explicitly visualise all the atom coordination within the unit cell. The electron density distribution maps of $Li_2Ni_2TeO_6$ along with the projected density of states (PDOS) plots have been availed in **Figures S7** and **S8**.

The observations made from the optimised $Li_2Ni_2TeO_6$ structure bear significance in rechargeable battery applications after recent reports deemed the layered material to be a propitious high-voltage cathode material. However, $Li_2Ni_2TeO_6$ was found to adopt various polytypes wherein different coordination between Li and oxygen emerge, depending on the synthesis technique employed.[20] Due to the comparable Shannon-Prewitt ionic radii of Li (0.76 Å in $LiO_6$ octahedra) and Ni (0.69 Å in $NiO_6$ octahedra),[61] certain preparation techniques for instance, high-temperature solid-state ceramics route compel the two atom species to swap their crystallographic positions (a phenomenon referred to as 'cationic mixing'), thus creating a disordered crystal framework of $Li_2Ni_2TeO_6$. The occupancy of Ni atoms in the same plane as Li atoms in the disordered $Li_2Ni_2TeO_6$ framework, significantly inhibits Li diffusion in the alkali atom plane thereby resulting in poor electrochemical performance.[20] As a strategy to improve the electrochemical performance of Li-based honeycomb layered oxides, an ordered $Li_2Ni_2TeO_6$ (with varied crystal versatility) can be designed via topochemical ion-exchange (metathesis reaction), which involves a low-temperature heat-treatment of $Na_2Ni_2TeO_6$ alongside a molten Li salt such as $LiNO_3$.[20] In this context, the present computational results evince the possibility of synthesising ordered $Li_2Ni_2TeO_6$ polytypes also from $K_2Ni_2TeO_6$ via topochemical ion-exchange.

### 2.2. Crystal structure of $H_2Ni_2TeO_6$

Layered frameworks that can particularly accommodate hydrogen atoms are relatively limited due to their complex syntheses and their highly unstable structures that hinder experimental observations. In fact, layered oxides encompassing hydrogen atoms sandwiched between transition metal slabs have not been reported to date. Nonetheless, we show here that the experimental limitations can be circumvented via DFT computations which aptly predict the optimised structure of $H_2Ni_2TeO_6$ shown in **Figures 6a** and **6b**. The optimised lattice parameters have been furnished as **Table S3**. The $H_2Ni_2TeO_6$ crystal structure shows the hydrogen atoms to be linearly coordinated with two oxygen atoms from the adjacent transition metal slabs forming dumbbell-like conformations. However, the hydrogen atom was observed to preferentially form a stronger bond (closer coordination) with one of the oxygen atoms creating the hydroxyl (OH$^-$) group.

In contrast with the other honeycomb layered tellurate compositions modelled in this study, the most stable structure of $H_2Ni_2TeO_6$ displays a ferromagnetic state of transition metal atoms (*albeit* antiferromagnetic states were also observed), as illustrated in **Table S3**. The high stability of the ferromagnetic state is attributed to the decrease in super-exchange interactions that typically accompany the formation of OH$^-$ groups.[62,63] These results implies that moisture absorption by alkali cations will change the magnetic properties of $[Ni_2TeO_6]^{2-}$. Further, the theoretical predictions made in this study pave way for future experimental design and synthesis of $H_2Ni_2TeO_6$ which is envisaged to find niche utility in the realms of solid-state hydrogen storage and catalysis.

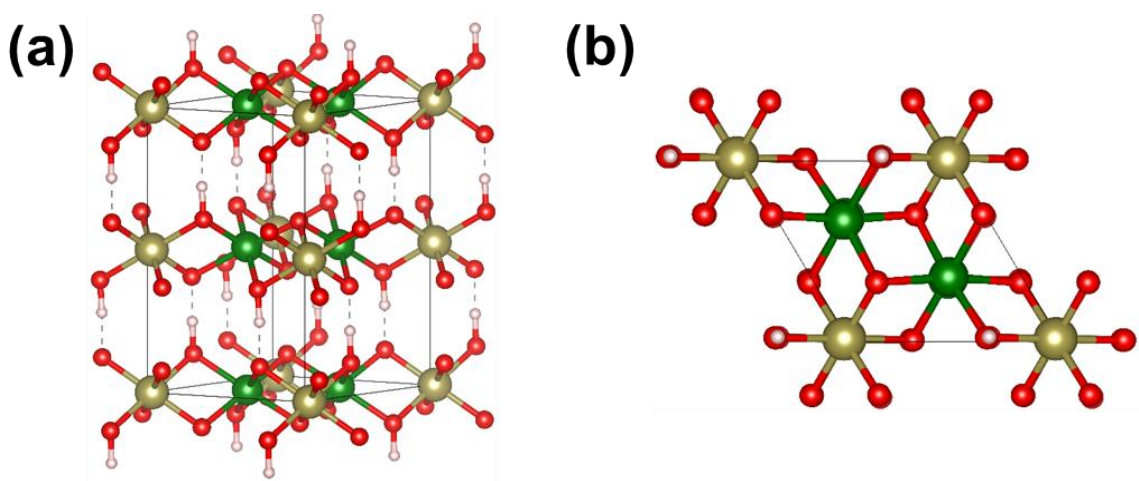

**Figure 6. Optimised crystal structural frameworks of $H_2Ni_2TeO_6$.** Optimised crystal structural framework of $H_2Ni_2TeO_6$ when viewed along **(a)** [110] zone axis and **(b)** [001] zone axis. Hydrogen atoms are shown in rose pink. Ni atoms (dark green) are arranged in a honeycomb configuration around Te atoms (in ochre) via the oxygen atoms (in red) to form $NiO_6$ and $TeO_6$ octahedra. Note that the crystal structural framework is slightly deviated from the [110] zone axis, in order to explicitly visualise all the atom coordination within the unit cell.

## 2.3. Crystal structures of $Ag_2Ni_2TeO_6$, $Au_2Ni_2TeO_6$ and $Cu_2Ni_2TeO_6$

The crystal versatility of honeycomb layered oxides, evidenced by their ability to accommodate a variety of atom species within the lattice structures, is a critical distinguishing property of this class of layered oxides. This diversity has been attested by experimental reports on honeycomb layered oxides featuring coinage-metal atoms (such as Ag and Cu) between the honeycomb slabs. [30,64-67] Replacing the alkali atoms with coinage-metal atoms is posited to ameliorate the functionalities of these materials, thereby expanding their range of utility in applications such as catalysis, magnetism, and optics. Therefore, for insight into the stability of these emergent structures, the $K_2Ni_2TeO_6$ crystal lattice was investigated with Cu, Ag, and Au coinage-metal atoms. Group 11 elements (which consist of Cu, Ag, Au, amongst others) are *s*-orbital systems with fully closed *d*-valence electrons. However, these elements are also transition metals and are thus expected to undergo ionisation, leading to the formation of compounds manifesting *sd* hybridisation. Therefore, it is expected that the interlayer distance does not scale with Shannon-Prewitt ionic radii.

**Figures 7a** and **7b** show the optimised crystal structure of $Cu_2Ni_2TeO_6$ viewed along various zone axes. The optimised lattice parameters for $Cu_2Ni_2TeO_6$, along with those of $Ag_2Ni_2TeO_6$, and $Au_2Ni_2TeO_6$ are provided in **Table S2**. The DFT model shows Cu atoms to be linearly coordinated with two oxygen atoms, forming dumbbell-like conformations. Similar coordinations have also been experimentally observed in Cu-containing Delafossites such as $Cu_3Co_2SbO_6$ and $Cu_3Ni_2SbO_6$, [65,66] which signals the possibility of attaining tellurate composition analogues via experimental routes. Likewise, the $Ag_2Ni_2TeO_6$ optimised crystal structure (**Figures 7c** and **7d**) exhibited dumbbell-like coordinations resembling those in the $Cu_2Ni_2TeO_6$ structure —indicating their structural comparability. The coordination between the Ag and oxygen atoms predicted here has also been reported by experimental studies on other silver-based honeycomb layered oxides such as the antimonates ($Ag_3M_2SbO_6$ ($M$ = Ni, Co, Zn))[67] and bismuthates ($Ag_3Ni_2BiO_6$).[64] Other metastable configurations of $Cu_2Ni_2TeO_6$ and $Ag_2Ni_2TeO_6$ that show different coordination have been predicted in the present study (**Figures S10** and **S11**), wherein Ag is octahedrally coordinated to six oxygen atoms in a prismatic fashion (akin to the coordination exemplified in $AgRuO_3$[68]), whilst Cu is coordinated to four oxygen atoms forming $CuO_4$ plaquettes (as those observed in cuprates such as $Li_2CuO_2$[69-70]). The electron density distribution maps along with the projected density of states (PDOS) plots for the metastable configurations of $Cu_2Ni_2TeO_6$ and $Ag_2Ni_2TeO_6$ have been provided in **Figures S12** and **S13**, respectively. Unlike Ag and Au, the tendency for Cu atoms to also coordinate with four oxygen atoms (as predicted by DFT) can be ascribed to a stronger *d*-orbital hybridisation of Cu. Although these configurations are predicted to be metastable by the present DFT study, high-pressure synthesis can be a route to access honeycomb layered oxides with such atom coordination as pressure can induce a change in the bonding coordination of compounds. [1]

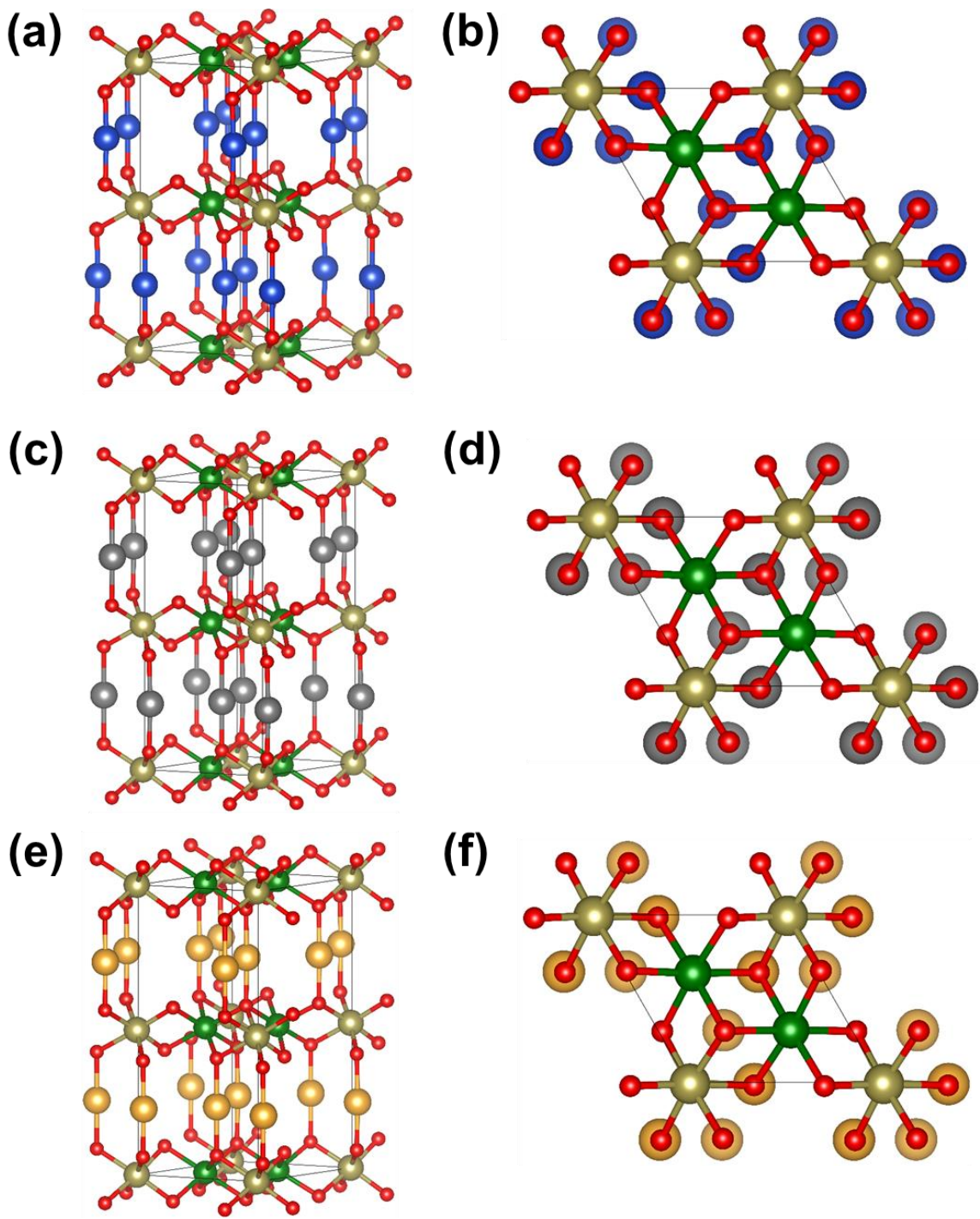

**Figure 7. Optimised crystal structural frameworks of $A_2Ni_2TeO_6$ ($A$ = Cu, Ag and Au).** Optimised crystal structural framework of $Cu_2Ni_2TeO_6$ when viewed along **(a)** [110] zone axis and **(b)** [001] zone axis. Copper atoms are shown in blue. **(c)** Optimised crystal structural framework of $Ag_2Ni_2TeO_6$ when viewed along the [110] zone axis and **(d)** when viewed along [001] zone axis. Silver atoms are shown in grey. **(e)** Optimised crystal structural framework of $Au_2Ni_2TeO_6$ when viewed along the [110] zone axis and **(f)** when viewed along [001] zone axis. Gold atoms are shown in orange-brown. Ni atoms (dark green) are arranged in a honeycomb configuration around Te atoms (in ochre) via the oxygen atoms to form $NiO_6$ and $TeO_6$ octahedra. Note that the crystal structural framework is slightly deviated from the [110] zone axis, in order to explicitly visualise all the atom coordination within the unit cell. The PDOS plots for these stable structures have been provided in **Figure S9**.

The predominance of linear coordination amongst coinage-metal atoms with oxygen atoms is attributed to the hybridised $d_z^2$ orbitals of the group 11 monovalent cations ($Cu^+$, $Ag^+$, $Au^+$, *etc*.). Distinct from what was observed in alkali metal atoms (**Figures S1**, **S2, S3, S4, S5, S6, S7** and **S8**), the valence electron distribution maps (shown in **Figures 8a** and **8b**) explicitly show the hybridisation of $d_z^2$ orbitals of coinage metal atoms with the $p_z$ orbitals of O. This can further be confirmed from the PDOS plots (**Figure S9**), showing valence band states emanating from the *d*-orbitals. **Figures S12** and **S13** show the valence electron distribution maps and PDOS plots of the metastable structures of $Cu_2Ni_2TeO_6$ and $Ag_2Ni_2TeO_6$, respectively. Although the valence band states emanating from the *d*-orbitals can be observed akin to those in the most stable configurations, the orbital interaction (hybridisation) between the $p_z$ and $d_z^2$ orbitals is dominant in the most stable configurations. This inclination towards dumbbell-like linear coordinations instead of prismatic coordinations engenders larger interlayer distance in $Ag_2Ni_2TeO_6$ than its Na analogue: in congruence with previous experimental reports on Ag-based honeycomb layered oxides, which further validate the optimised $Ag_2Ni_2TeO_6$ structure. [49,52] Considering that the detailed experimental synthesis and structural characterisation of $Ag_2Ni_2TeO_6$ has not yet been reported to date, [31] the results of this study will be instrumental in predicting emergent features of this material that we are currently pursuing. However, it is worth noting that given the unique chemistry of silver, we can anticipate a diverse assortment of atomic coordinations in Ag$_2$Ni$_2$TeO$_6$, not necessarily predicted by the DFT calculations in the present work.

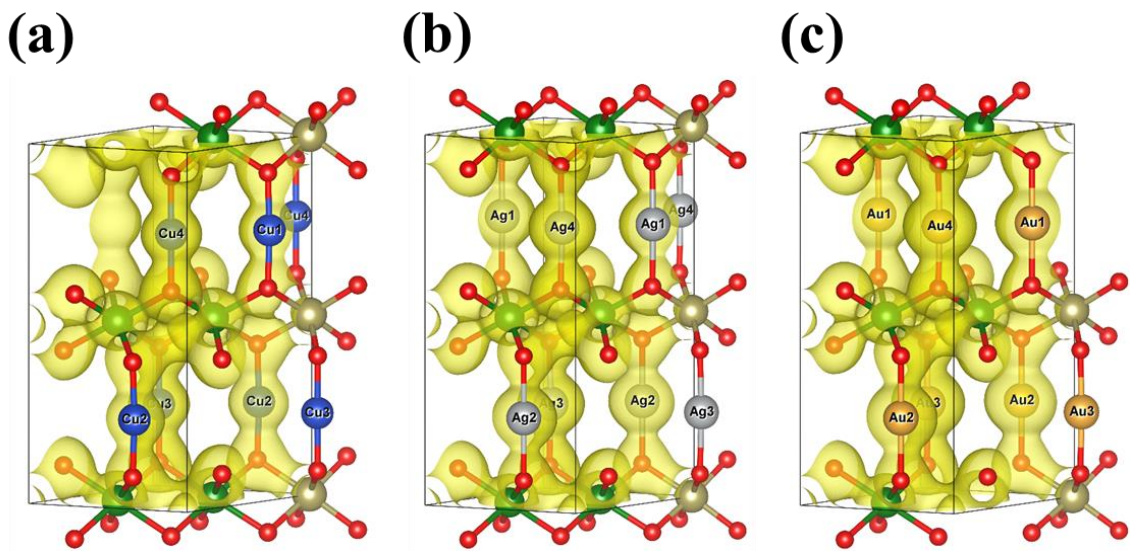

**Figure 8. Electron density distribution maps (yellow) of the most stable structures of *A*$_2$Ni$_2$TeO$_6$ (*A* = Cu, Ag, Au).** Electron density distribution map of **(a)** Cu$_2$Ni$_2$TeO$_6$ (for the structure shown in **Figures 7a** and **7b**), **(b)** Ag$_2$Ni$_2$TeO$_6$ (for the structure shown in **Figures 7c** and **7d**) and **(c)** Au$_2$Ni$_2$TeO$_6$ (for the structure shown in **Figures 7e** and **7f**).

Nonetheless, the ability to predict novel material compositions—such as H$_2$Ni$_2$TeO$_6$ and Ag$_2$Ni$_2$TeO$_6$—that may go beyond the boundaries of present experimental validation earmarks the expedience of DFT computation methods in quest for novel honeycomb layered oxides. Amongst the coinage-metal based materials investigated, Au-related compounds are far less investigated than the other group 11 element species such as Cu and Ag. In fact, no Au-based layered oxides have been reported to date. Au-based compounds have been found to manifest high activity and selectivity for many reactions—features that are particularly attractive in the fields of catalysis and optics.[86,87] As such, the stability of Au atoms in the K$_2$Ni$_2$TeO$_6$ lattice framework was investigated, as shown in **Figures 7e** and **7f**. Akin to Cu$_2$Ni$_2$TeO$_6$ and Ag$_2$Ni$_2$TeO$_6$, Au$_2$Ni$_2$TeO$_6$ exhibits a linear coordination of Au atoms with two oxygen atoms in a dumbbell-like conformation. As expected, the Au atoms render a large interlayer distance in the linearly coordinated structure. Additionally, the valence electron distribution for Au$_2$Ni$_2$TeO$_6$ (**Figure 8c**) is similar to that of Cu$_2$Ni$_2$TeO$_6$ and Ag$_2$Ni$_2$TeO$_6$, affirming interaction between the $d_z^2$ orbitals of Au and $p_z$ orbitals of O. The PDOS of Au$_2$Ni$_2$TeO$_6$ is also similar to that of Ag$_2$Ni$_2$TeO$_6$ and Cu$_2$Ni$_2$TeO$_6$ (**Figure S9**), except that the hybridisation of O and Au bands lying at the vicinity of the Fermi energy is small. Nevertheless, what distinguishes $Au_2Ni_2TeO_6$ from $Cu_2Ni_2TeO_6$ and $Ag_2Ni_2TeO_6$ is that the peaks arising from Ni and O hybridisation cannot be observed at a range of 4 eV below the Fermi energy. As for $Cu_2Ni_2TeO_6$ and $Ag_2Ni_2TeO_6$, the peaks are located at 2.5 eV and 0.7 eV below the Fermi energy, respectively. This implies that the physicochemical properties of $Au_2Ni_2TeO_6$ are most likely affected by Au-derived orbitals, envisaging Au–O–Ni interactions. Given the untapped state of this research space, the Au-based honeycomb layered oxide presented here promises to rekindle interest in the realm of noble gold chemistry since the synthesis and evaluation of $Au_2Ni_2TeO_6$ herald major advancements in the exploration of the piquant catalytic and optical properties of small gold clusters, mono-atomically dispersed gold, and monovalent gold complexes.

## 3. <u>IMPLICATIONS</u>

### 3.1. Interlayer distances of $A_2Ni_2TeO_6$ ($A$ = H, Li, Na, Rb, Cs, Ag, Au and Cu)

As shown in **Figure 9**, the $A_2Ni_2TeO_6$ ($A$ = H, Li, Na, Rb, Cs, Ag, Au, and Cu) honeycomb tellurates presented in this study exhibit distinct structures with varying interslab/interlayer distances depending on the species of $A^+$ cations employed. We note here that the straight-line correlation between the interslab distance and the Shannon-Prewitt ionic radius size of the resident $A^+$ cations is only valid for the alkali cations, *i.e.,* Li, Na, Ag (prismatic), K, Rb, and Cs. Consequently, other metastable honeycomb layered tellurates with polyhedral coordinations that do not exhibit this trend have also been found. In particular, Li atoms and oxygen form octahedral coordination, whilst Ag and Cu form a linear/dumbbell coordination with oxygen, thus differing from the rest that are prismatic. This is because alkali atoms have high ionicity (thus lower covalency between alkali $A$ and O). Moreover, Cu also has an additional metastable coordination with oxygen that is square planar. Generally, a small ionic radius favours octahedral, square planar and/or dumbbell/linear coordination with a smaller or intermediate interlayer distance whereas a larger radius predominantly favours a prismatic coordination with the largest interlayer distances. In particular, with regards to coinage-metal atoms (*viz*., Cu, Ag, and Au), the linear coordination between $A^+$ cations and the oxygen atoms engenders intermediate interslab distances even with smaller $A^+$ cations. For instance, even though the Shannon-Prewitt ionic radius of monovalent Cu is comparable to that of Li, the interslab distance in the $Cu_2Ni_2TeO_6$ crystal structure is still larger than that of $Li_2Ni_2TeO_6$. This is ascribed to the covalent bonds between *A* and O arising due to the hybridisation of *d* orbitals and O–2*p* orbitals.

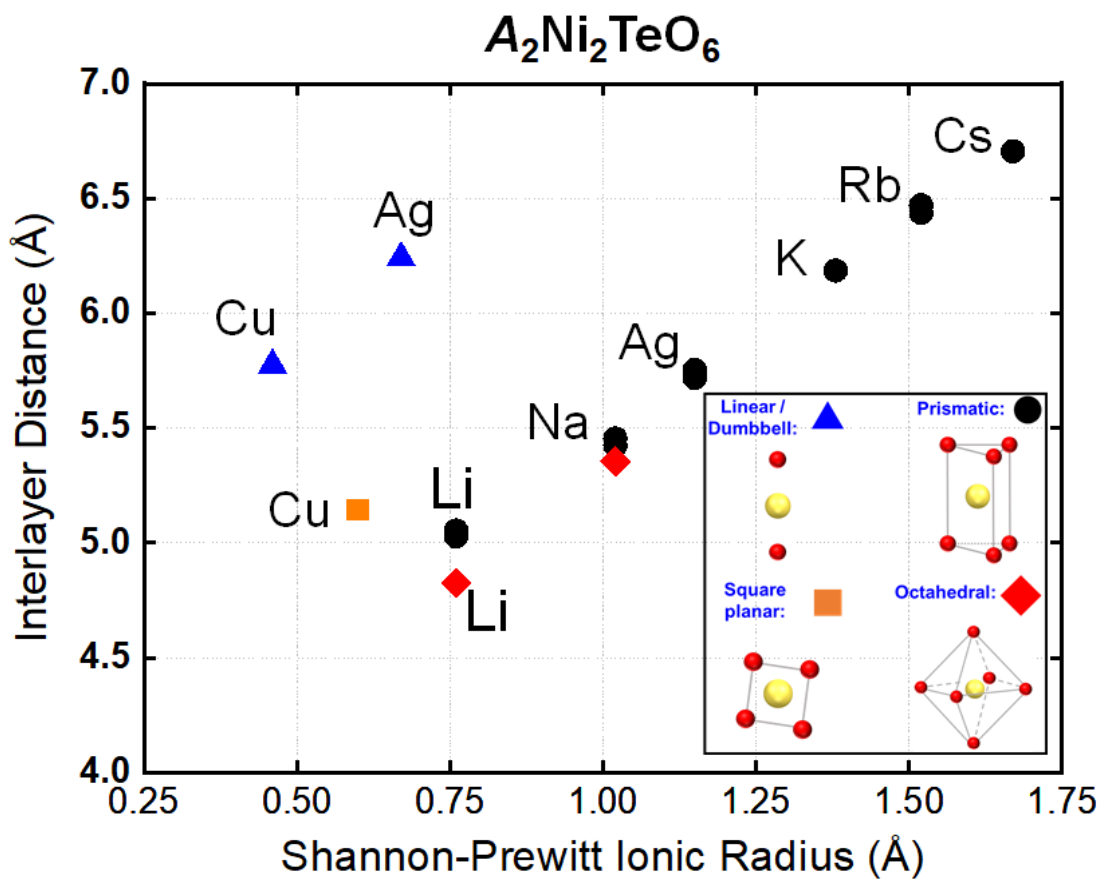

**Figure 9. Interslab/interlayer distances rendered by the $A_2Ni_2TeO_6$ ($A$ = Li, Na, Cu, Ag, K, Rb and Cs) compositions.** The changes in the interslab distances in the $A_2Ni_2TeO_6$ compositions are derived using first-principles density functional theory calculations. The Shannon[58] radii of the monovalent $A^+$ cations is shown in the horizontal scale. Note that the Shannon ionic radius value of Au when coordinated to two oxygen atoms is not available; thus has not been shown in the plots along with that of H. Note that, each of the prismatic coordination plots for $Na_2Ni_2TeO_6$, $Ag_2Ni_2TeO_6$, $Rb_2Ni_2TeO_6$ and $Cs_2Ni_2TeO_6$ include two values. The interslab distance corresponds to the distance between transition metal atoms lying directly above or below the adjacent layers.

The interslab distances of the honeycomb layered tellurates presented in this study significantly influence emergent functionalities such as solid-state cationic diffusion. [1,71] In fact, a recent study highlighted the correlation between the coordination structure, the size of the $A^+$ alkali cations, and the resulting ionic conductivity of honeycomb layered oxides. [1] For instance, the small ionic radius of $Li^+$ begets strongly bonded octahedral structures with smaller inter-layer distances that result in limited ionic conductivity. However, the larger ionic radius of $K^+$ yields weakly-bonded prismatic structures with larger inter-layer distances and enhanced ionic conductivities.

On the other hand, the linear coordination between coinage-metal cations such as $Ag^+$ and oxygen atoms precipitates relatively weaker interlayer bonds and consequently intermediate inter-layer distances in the pertinent structures. In principle, the weaker interlayer bonds (and intermediate and large interlayer distances) in linear or prismatic layered structures are expected to create more open voids within the transition metal layers/slabs allowing for facile two-dimensional diffusion of alkali or coinage-metal atoms within the slabs. In this vein, pursuing the design of new honeycomb compositions predicted in this study, such as $Rb_2Ni_2TeO_6$, $Cs_2Ni_2TeO_6$, and $Ag_2Ni_2TeO_6$, can therefore be a worthwhile undertaking in fields such as materials science geared towards various functionalities such as magnetism, catalysis and energy storage solutions.

*Nota bene*, honeycomb layered oxides display a multitude of slab stacking structures typically characterised by the Hagenmuller-Delmas notation. [72] In this study, the $K_2Ni_2TeO_6$ used as the initial model structure displays a *prismatic* coordination of K atoms with *two* repetitive honeycomb layers (comprising $NiO_6$ and $TeO_6$ octahedra) for each unit cell (P2-Type in Hagenmuller-Delmas notation). [23,73] As such, the honeycomb layered tellurate compositions predicted in this study also show P2-Type structural configurations. This suggests that the new honeycomb layered oxides can be experimentally reproduced via topotactic ion-exchange. Previous reports have also shown layered oxide structures manifest a plenitude of stacking faults and other topological defects that bring forth fascinating functionalities. Therefore, we should expect the materials predicted herein to exhibit various stacking faults and other topological defects accompanied by unique properties. [1,74,75] On a separate note, the coordination chemistry of coinage-metal atoms such as Ag can be expected to show other variegated coordinations besides the typical linear coordination envisioned in this study. [1,64,67] This is because some coinage-metal atoms, such as Ag in $Ag_2NiO_2$, have been known to exhibit unprecedented sub-valency states (*i.e.*, valency states between +1 and 0) [76] that lead to idiosyncratic structural and bonding properties when sandwiched between transition metal layers. Therefore, even though first-principles predictions vaticinate the design and prospects of new honeycomb tellurate compositions such as $Ag_2Ni_2TeO_6$ and $Au_2Ni_2TeO_6$, experimental inquests are still vital to advancing this class of materials.

## 3.2. Cation honeycomb lattice and the structural diversity of coinage-metal atom layered tellurates

In all the studied honeycomb layered tellurates, the cations form a honeycomb lattice where the unit cell consists of two individual *A* cations, as shown in **Figure 10**. (The interatomic distances of the cations in the honeycomb lattice are provided in **Table S4**). Thus, each cation belongs to a particular triangular sub-lattice of the honeycomb cation lattice, identified by a degree of freedom known as a pseudo-spin as shown in **Figure 10 (a)**. In fact, the honeycomb carbon lattice structure of graphene requires such an additional degree of freedom to describe the orbital wave functions of conduction electrons sitting in each sub-lattice.[85] Meanwhile, it has been previously argued that the honeycomb lattice is a unique exemplar of crystalline symmetries known as modular/conformal invariance, namely scale, translation and rotation symmetries.[86] Conformal symmetries in honeycomb layered tellurates are particularly powerful since they not only completely fix the geometry, topology and diffusion dynamics of the cations within the material, but also predict new meta-stable structures via phase-transition when spontaneously broken. This symmetry breaking is interpreted as the mutual attraction of the cations in each sub-lattice with opposite pseudo-spin in each unit cell, corresponding to a metallophilic interaction.[80-82]

In the exemplar case of Ag, preliminary experimental results show a pair of triangular lattices of cations stacked on each other, whereby the honeycomb cation lattice, shown in **Figure 10 (b),** appears to bifurcate into two triangular sub-lattices as shown in **Figure 10 (c)**, whereby the metallophilic interactions between the pairs of opposite pseudo-spin cations (indicated by the blue double arrows), manifest as a pseudo-magnetic field acting on the pseudo-spins.[79] However, in our DFT simulation results, we only found the metastable structure of $Ag_2Ni_2TeO_6$ with a linear/dumbbell coordination and the cation honeycomb lattice structure intact. This can be explained by

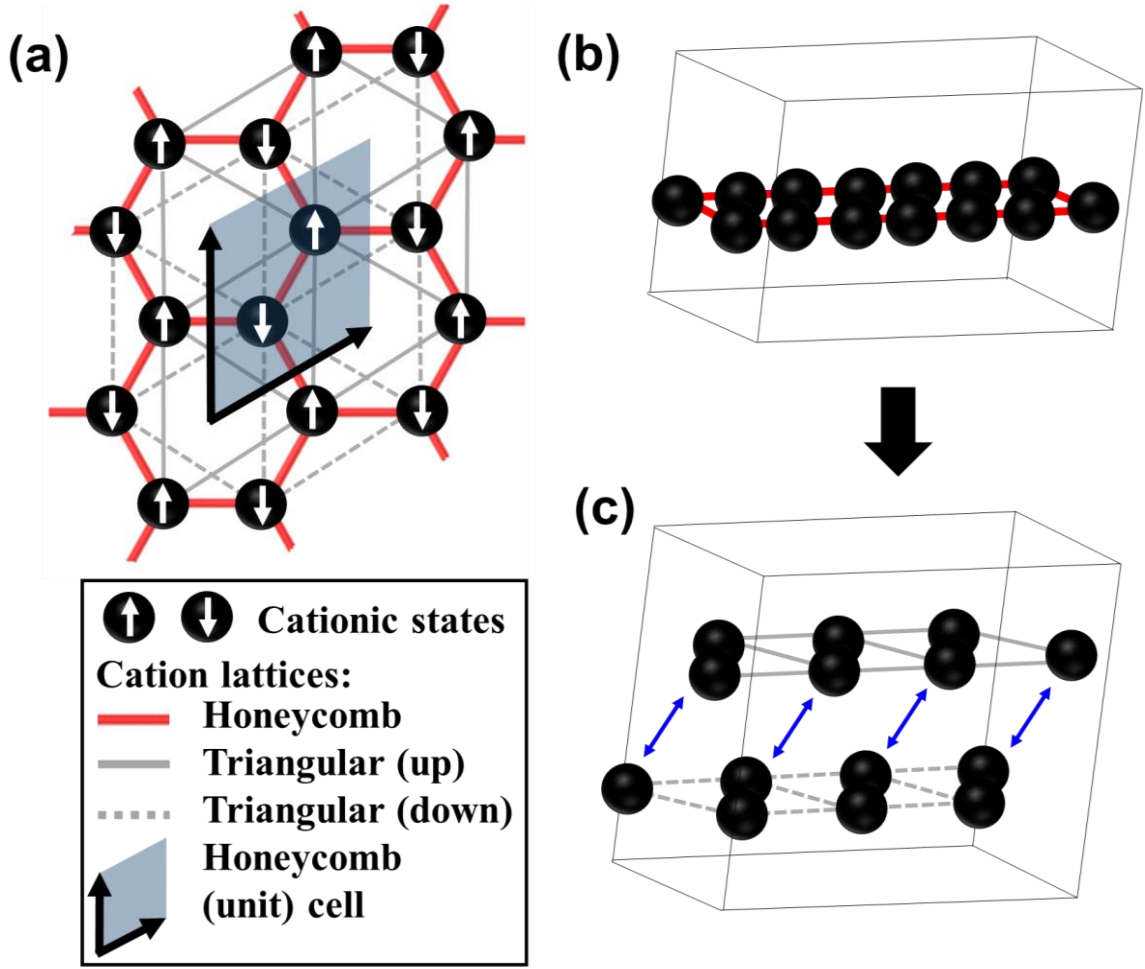

**Figure 10: Bifurcation of the cation honeycomb lattice of coinage metal atoms into a pair of triangular sub-lattices stabilized by metallophilic interactions.** (a) The honeycomb lattice (black spheres connected by red lines) of the $A$ = Li, Na, K, Cu, Ag, Au, Cs and Rb cations in $A_2Ni_2TeO_6$, viewed in the [001] direction, as found by our DFT simulation. The cation honeycomb lattice is comprised of a pair of up and down triangular sub-lattices (grey solid and dashed lines respectively), assigning opposite pseudo-spins (indicated by white arrows) to each cation in the unit cell (grey rhombus bounded by black unit vectors). (b) A perspective view of the unit cell of $A_2Ni_2TeO_6$ shows the alignment of the cations into a honeycomb pattern. (c) The honeycomb lattice of coinage metal atoms ($A$ = Cu, Ag, Au) bifurcates into a bilayer of opposite pseudo-spin triangular lattices, whereby metallophilic interactions acting as a pseudo-magnetic field between the opposite pseudo-spin coinage metal atoms, as

<span style="color: blue">indicated by the blue double arrows, stabilise the bilayer.</span>

the fact that the Kohn-Sham formalism employed in our DFT calculations is not a straightforward procedure to effectively treat the metallophilic interactions which are known to be non-negligible particularly for coinage-metal atoms. [80-82] The particular trend in **Figure 9,** that coinage metals form a linear/dumbbell coordination with oxygen in the studied layered tellurates suggest that the reported bifurcation of the honeycomb lattice can be theoretically attributed to the unique geometric features of the honeycomb lattice of cations influenced by particular polyhedral coordination of Ag with oxygen, metallophilicity and possibly sub-valency. Specifically, the breaking of the aforementioned conformal symmetry, whereby the requisite conditions for the monolayer-bilayer phase transition can be attributed to a dumbbell/linear coordination coupled with the presence of finite argentophilic attractions which stabilises the bilayer. [79] At the critical point where the cations are in a honeycomb lattice (corresponding to the simulation results herein), conformal symmetry is restored, and the theory of cationic diffusion and vacancies corresponds to a dual gravity theory in hyperbolic geometry known as Anti-de Sitter space.[79,88]

## 3. CONCLUSION

This study aims to extend the frontier of nickel-based tellurates and diversify their functionalities by employing density functional theory (DFT) to explore the structural configuration of new honeycomb layered tellurates entailing previously unreported cation species such as large alkali atoms (in $Rb_2Ni_2TeO_6$ and $Cs_2Ni_2TeO_6$), coinage-metal atoms (in $Cu_2Ni_2TeO_6$, $Ag_2Ni_2TeO_6$, and $Au_2Ni_2TeO_6$) and hydrogen (in $H_2Ni_2TeO_6$). The theoretical simulations, employing $K_2Ni_2TeO_6$ as the initial structural model, reveal an underlying correlation between the distinct coordination structures engendered by the different species (in this case: large alkali atoms and coinage metal atoms) and their resultant interslab / interlayer distances. The interlayer distance in honeycomb layered tellurates comprising large alkali atoms (Rb and Cs) is seen to be contingent on the ionic radius of the atom, which forms a prismatic coordination with the oxygen atoms from the adjacent honeycomb slabs. On the other hand, coinage-metal atoms (Cu, Ag, and Au) show a proclivity towards linear coordination with two oxygen atoms rendering intermediate interlayer distances independent of their ionic radii. Further, the hydrogen atom is seen to render a linear coordination with two oxygen atoms whilst manifesting a preferential coordination with one of the oxygen atoms that results in the formation of hydroxyl groups. Indeed, honeycomb layered oxides such as $K_2Ni_2TeO_6$ are hygroscopic;[23] thus, capable of forming hydroxyl groups upon absorption of moisture. The formation of hydroxyl group results in a change in magnetic configuration (antiferromagnetic to ferromagnetic transition) of the honeycomb slab, envisaging applications in magnetic switching.

The new honeycomb layered compositions presented herein betoken promising functionalities such as two-dimensional magnetism, optics and catalysis for utility in fields ranging from solid-state hydrogen storage to noble gold chemistry. However, even with the excellent prospects envisioned in this study, experimental techniques remain a requisite in the design and exploration of the new materials due to the possibility of disparate emergent properties such as the variegated bonding properties of coinage metals such as Ag atoms. By considering the known chemistry of 1) metallophilicity and possible sub-valencies of coinage-metal atoms [76,77,80-82]; our computational results imply that the found configurations of the coinage metal oxide tellurates satisfying the additional conditions: 2) a honeycomb lattice of cations; and 3) a linear/dumbbell coordination of cations with oxygen atoms; harbour the possibility for the observation of a monolayer-bilayer phase transition, with crucial implications for the multi-disciplinary field of analogue condensed matter systems of gravity. [1,43-45,88] Thus, the present findings not only envisage the vast compositional space of honeycomb layered tellurates but also paves the way future explorations and the rational design of new-fangled honeycomb layered oxides with assorted physicochemical properties.

# SUPPLEMENTARY TABLES AND FIGURES

**Table S1. Optimised lattice parameters of $K_2Ni_2TeO_6$.** The magnetic atoms (Ni) were assessed in two configurations: ferromagnetic (FM) and antiferromagnetic (AFM) states. The experimentally obtained lattice parameters are shown for a comparison.

| Compound | State | $a$ (Å) | $b$ (Å) | $c$ (Å) | $\alpha$ ° | $\beta$ ° | $\gamma$ ° |
|---|---|---|---|---|---|---|---|
| $K_2Ni_2TeO_6$ | AFM | 5.28176 | 5.28176 | 12.37282 | 90.0000 | 90.0000 | 120.0588 |
|  | FM | 5.32141 | 5.32141 | 12.49672 | 90.0000 | 90.0000 | 120.1055 |
| Experimental value |  | 5.26060 | 5.26060 | 12.46690 | 90.0000 | 90.0000 | 120.0000 |

**Table S2. Optimised lattice parameters for various configurations of $A_2Ni_2TeO_6$ ($A$ = Li, Na, Rb, Cs, Cu, Ag, and Au).** The magnetic atoms (Ni) were assessed in antiferromagnetic (AFM) states.

| Compound | Structure | $a$ (Å) | $b$ (Å) | $c$ (Å) | $\alpha$ ° | $\beta$ ° | $\gamma$ ° |
|---|---|---|---|---|---|---|---|
| $Na_2Ni_2TeO_6$ | Fig. 2a | 5.21403 | 5.21403 | 10.85093 | 90.3336 | 89.6702 | 120.2777 |
|  | Fig. 2c | 5.17381 | 5.17454 | 10.91146 | 90.0022 | 89.9980 | 119.9526 |
|  | Fig. 2e | 5.17516 | 5.17554 | 10.71164 | 90.8184 | 89.1720 | 119.4091 |
| $Rb_2Ni_2TeO_6$ | Fig. 3a | 5.33225 | 5.33225 | 12.94207 | 89.9998 | 89.9999 | 120.0310 |
|  | Fig. 3c | 5.31938 | 5.31932 | 12.87379 | 90.0074 | 89.9947 | 120.0477 |
| $Cs_2Ni_2TeO_6$ | Fig. 4a | 5.42700 | 5.42700 | 13.45493 | 90.0001 | 90.0001 | 119.9957 |
|  | Fig. 4c | 5.40944 | 5.40944 | 13.33245 | 90.0071 | 89.9920 | 119.9952 |
| $Li_2Ni_2TeO_6$ | Fig. 5a | 5.16340 | 5.16355 | 10.05834 | 90.1412 | 89.8514 | 120.5663 |
|  | Fig. 5c | 5.11048 | 5.11045 | 10.10848 | 90.0109 | 89.9975 | 119.9924 |
|  | Fig. 5e | 5.13412 | 5.13638 | 9.65122 | 90.0785 | 89.9220 | 119.5745 |
| $Cu_2Ni_2TeO_6$ | Fig. 6a | 5.24766 | 5.24777 | 11.55000 | 90.0468 | 89.9602 | 119.8559 |
| $Ag_2Ni_2TeO_6$ | Fig. 6c | 5.23096 | 5.23123 | 12.47119 | 90.0076 | 89.9901 | 119.8416 |
| $Au_2Ni_2TeO_6$ | Fig. 6e | 5.25312 | 5.25337 | 12.31434 | 90.0482 | 89.9593 | 119.6811 |

**Table S3. Optimised lattice parameters for various configurations of $H_2Ni_2TeO_6$.** The magnetic atoms (Ni) were assessed in two configurations: ferromagnetic (FM) and antiferromagnetic (AFM) states. The lattice parameters for the most stable configuration are highlighted in grey.

| Compound | State | $a$ (Å) | $b$ (Å) | $c$ (Å) | $\alpha$ ° | $\beta$ ° | $\gamma$ ° |
|---|---|---|---|---|---|---|---|
| $H_2Ni_2TeO_6$ | AFM-1 | 5.26335 | 5.26356 | 9.18956 | 95.7798 | 84.2228 | 119.7717 |
| | AFM-2 | 5.26488 | 5.26482 | 9.18296 | 96.0979 | 83.9147 | 119.7702 |
| | AFM-3 | 5.26714 | 5.26683 | 9.18887 | 95.9650 | 84.0158 | 119.7788 |
| | FM-1 | 5.26230 | 5.26238 | 9.28724 | 90.0835 | 89.9207 | 120.4790 |
| | FM-2 | 5.28817 | 5.28811 | 9.23936 | 90.4659 | 89.5393 | 120.5527 |
| | FM-3 | 5.29236 | 5.29344 | 9.28540 | 90.3051 | 89.7239 | 120.5348 |

**Table S4. Interatomic distances of alkali- or coinage-metal atoms aligned in a honeycomb lattice in various configurations of $A_2Ni_2TeO_6$ (A=Li,Na,K,Rb,Cs,Ag,Cu and Au).**

| Compound | Alkali/Coinage Metal Coordination | Alkali/Coinage Metal Interatomic Distances (Å) |
|---|---|---|
| $Li_2Ni_2TeO_6$ | Octahedral | 2.97589, 2.96773, 2.97002 |
| $Li_2Ni_2TeO_6$ | Prismatic | 2.89596, 3.01132, 3.01162 |
| $Li_2Ni_2TeO_6$ | Prismatic | 2.95115, 2.95230, 2.94847 |
| $Na_2Ni_2TeO_6$ | Octahedral | 2.95641, 2.96169, 3.07576 |
| $Na_2Ni_2TeO_6$ | Prismatic | 3.16920, 2.92141, 2.93557 |
| $Na_2Ni_2TeO_6$ | Prismatic | 2.98888, 2.98626, 2.98893 |
| $K_2Ni_2TeO_6$ | Prismatic | 2.98257, 2.98375, 3.18399 |
| $Rb_2Ni_2TeO_6$ | Prismatic | 3.14984, 3.04239, 3.04335 |
| $Rb_2Ni_2TeO_6$ | Prismatic | 3.07031, 3.07049, 3.07037 |
| $Cs_2Ni_2TeO_6$ | Prismatic | 3.11747, 3.11917, 3.16551 |
| $Cs_2Ni_2TeO_6$ | Prismatic | 3.12341, 3.12313, 3.12311 |
| $Ag_2Ni_2TeO_6$ | Linear | 3.06804, 3.03408, 2.96673 |
| $Ag_2Ni_2TeO_6$ | Prismatic | 2.90872, 2.92124, 3.23107 |
| $Ag_2Ni_2TeO_6$ | Prismatic | 3.01657, 3.03070, 3.00041 |
| $Cu_2Ni_2TeO_6$ | Linear | 3.08110, 3.18506, 2.84035 |
| $Cu_2Ni_2TeO_6$ | Square Planar | 2.83119, 2.98667, 3.27779 |
| $Au_2Ni_2TeO_6$ | Linear | 3.05699, 3.09101, 2.96739 |

**Table S5.** Local magnetic moment (in atomic units) of the atoms in the stable $A_2Ni_2TeO_6$ ($A$=K, Ag, Au and Cs) structures.

|  | $A$ = K | $A$ = Ag | | | $A$ = Au | $A$ = Cs | |
|---|---|---|---|---|---|---|---|
|  | Fig.1(a,b) | Fig.S7(a,b) | Fig.S7(c,d) | Fig.7(c,d) | Fig.7(e,f) | Fig.4(a,b) | Fig.4(c,d) |
| Te1 | 0.000 | 0.000 | 0.000 | −0.001 | 0.000 | 0.000 | 0.000 |
| Te2 | 0.000 | 0.000 | 0.000 | 0.001 | 0.000 | 0.000 | 0.000 |
| Ni1 | 1.699 | 1.685 | 1.676 | 1.693 | 1.681 | 1.698 | 1.695 |
| Ni2 | −1.699 | −1.685 | −1.676 | −1.694 | −1.682 | −1.698 | −1.695 |
| Ni3 | 1.699 | 1.685 | 1.676 | 1.694 | 1.682 | 1.698 | 1.695 |
| Ni4 | −1.699 | −1.685 | −1.676 | −1.693 | −1.681 | −1.698 | −1.695 |
| O1 | 0.001 | 0.004 | 0.000 | 0.003 | 0.000 | 0.000 | 0.000 |
| O2 | 0.001 | 0.004 | 0.000 | 0.003 | 0.000 | 0.000 | 0.000 |
| O3 | −0.001 | −0.005 | 0.000 | 0.001 | 0.000 | 0.000 | 0.000 |
| O4 | 0.000 | 0.002 | 0.000 | −0.005 | 0.000 | 0.000 | 0.000 |
| O5 | 0.000 | 0.002 | 0.000 | −0.005 | 0.000 | 0.000 | 0.000 |
| O6 | −0.001 | −0.005 | 0.000 | 0.001 | 0.000 | 0.000 | 0.000 |
| O7 | -0.001 | −0.004 | 0.000 | −0.003 | 0.000 | 0.000 | 0.000 |
| O8 | -0.001 | −0.004 | 0.000 | −0.003 | 0.000 | 0.000 | 0.000 |
| O9 | 0.000 | −0.002 | 0.000 | 0.005 | 0.000 | 0.000 | 0.000 |
| O10 | 0.001 | 0.005 | 0.000 | −0.001 | 0.000 | 0.000 | 0.000 |
| O11 | 0.001 | 0.005 | 0.000 | −0.001 | 0.000 | 0.000 | 0.000 |
| O12 | 0.000 | −0.002 | 0.000 | 0.005 | 0.000 | 0.000 | 0.000 |
| A1 | 0.000 | 0.000 | 0.000 | 0.000 | 0.000 | 0.000 | 0.000 |
| A2 | 0.000 | 0.000 | 0.000 | 0.000 | 0.000 | 0.000 | 0.000 |
| A3 | 0.000 | 0.000 | 0.000 | 0.000 | 0.000 | 0.000 | 0.000 |
| A4 | 0.000 | 0.000 | 0.000 | 0.000 | 0.000 | 0.000 | 0.000 |

**Table S6.** Local magnetic moment (in atomic units) of the atoms in the stable $A_2Ni_2TeO_6$ ($A$=Cu, Li, Na and Rb) structures.

|  | $A$ = Cu | | $A$ = Li | | | $A$ = Na | | | $A$ = Rb | |
|---|---|---|---|---|---|---|---|---|---|---|
|  | Fig.7(a,b) | Fig.S6(a,b) | Fig.5(a,b) | Fig.5(c,d) | Fig.5(e,f) | Fig.2(a,b) | Fig.2(c,d) | Fig.2(e,f) | Fig.3(a,b) | Fig.3(c,d) |
| Te1 | 0.000 | −0.001 | 0.000 | 0.000 | 0.000 | 0.000 | 0.000 | 0.000 | 0.000 | 0.000 |
| Te2 | 0.000 | 0.001 | 0.000 | 0.000 | 0.000 | 0.000 | 0.000 | 0.000 | 0.000 | 0.000 |
| Ni1 | 1.661 | 1.689 | 1.689 | 1.663 | 1.679 | 1.675 | 1.693 | 1.685 | 1.702 | 1.692 |
| Ni2 | −1.661 | −1.691 | −1.689 | −1.663 | −1.679 | −1.675 | −1.693 | −1.684 | −1.702 | −1.692 |
| Ni3 | 1.661 | 1.691 | 1.689 | 1.663 | 1.679 | 1.675 | 1.693 | 1.684 | 1.702 | 1.692 |
| Ni4 | −1.661 | −1.689 | −1.689 | −1.663 | −1.679 | −1.675 | −1.693 | −1.685 | −1.702 | −1.692 |
| O1 | −0.004 | 0.003 | 0.003 | 0.000 | 0.000 | 0.000 | 0.002 | 0.000 | 0.001 | 0.000 |
| O2 | −0.004 | 0.003 | 0.003 | 0.000 | 0.000 | 0.000 | 0.003 | 0.000 | 0.001 | 0.000 |
| O3 | −0.004 | 0.001 | −0.005 | 0.000 | 0.004 | 0.000 | −0.003 | 0.003 | −0.001 | 0.000 |
| O4 | 0.005 | −0.006 | 0.001 | 0.000 | −0.002 | 0.000 | 0.001 | −0.001 | −0.001 | 0.000 |
| O5 | 0.005 | −0.006 | 0.002 | 0.000 | −0.004 | 0.000 | 0.001 | −0.003 | −0.001 | 0.000 |
| O6 | −0.003 | 0.001 | −0.005 | 0.000 | 0.002 | 0.000 | −0.003 | 0.001 | −0.001 | 0.000 |
| O7 | 0.004 | −0.003 | −0.003 | 0.000 | 0.000 | 0.000 | −0.002 | 0.000 | −0.001 | 0.000 |
| O8 | 0.004 | −0.003 | −0.003 | 0.000 | 0.000 | 0.000 | −0.003 | 0.000 | −0.001 | 0.000 |
| O9 | −0.005 | 0.006 | −0.002 | 0.000 | 0.004 | 0.000 | −0.001 | 0.003 | 0.001 | 0.000 |
| O10 | 0.003 | −0.001 | 0.005 | 0.000 | −0.002 | 0.000 | 0.003 | −0.001 | 0.001 | 0.000 |
| O11 | 0.004 | −0.001 | 0.005 | 0.000 | −0.004 | 0.000 | 0.003 | −0.003 | 0.001 | 0.000 |
| O12 | −0.006 | 0.006 | −0.002 | 0.000 | 0.002 | 0.000 | −0.001 | 0.001 | 0.001 | 0.000 |
| A1 | 0.000 | 0.000 | 0.000 | 0.000 | −0.003 | 0.000 | 0.000 | −0.001 | 0.000 | 0.000 |
| A2 | 0.001 | 0.000 | 0.000 | 0.000 | 0.003 | 0.000 | 0.000 | 0.001 | 0.000 | 0.000 |
| A3 | 0.001 | 0.000 | 0.000 | 0.000 | −0.003 | 0.000 | 0.000 | −0.001 | 0.000 | 0.000 |
| A4 | 0.000 | 0.000 | 0.000 | 0.000 | 0.003 | 0.000 | 0.000 | 0.001 | 0.000 | 0.000 |

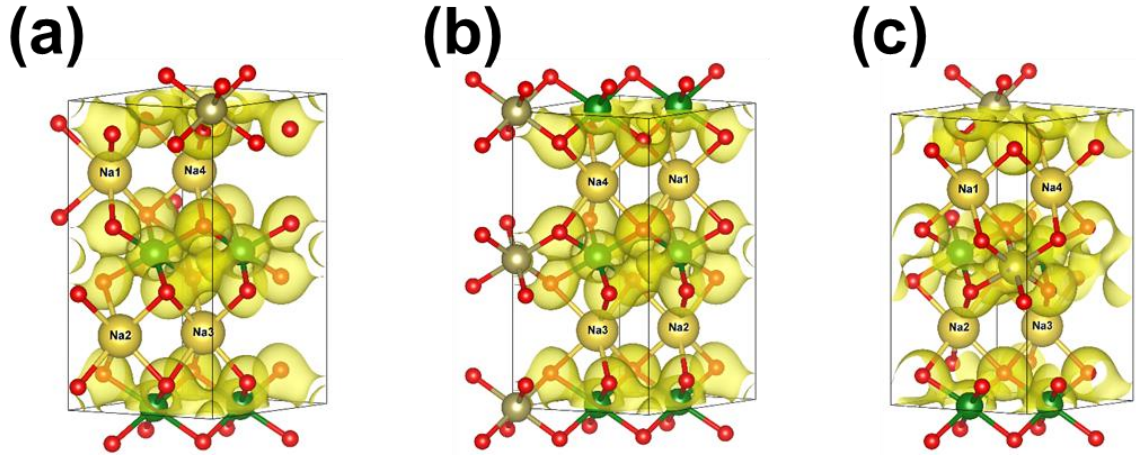

**Figure S1. Electron density distribution maps (yellow) of $Na_2Ni_2TeO_6$.** Electron density distribution maps of **(a)** $Na_2Ni_2TeO_6$ (for the structure shown in **Figures 2a** and **2b**), **(b)** $Na_2Ni_2TeO_6$ (for the structure shown in **Figures 2c** and **2d**) and **(c)** $Na_2Ni_2TeO_6$ (for the structure shown in **Figures 2e** and **2f**). The threshold is 0.05 e⁻/Bohr³.

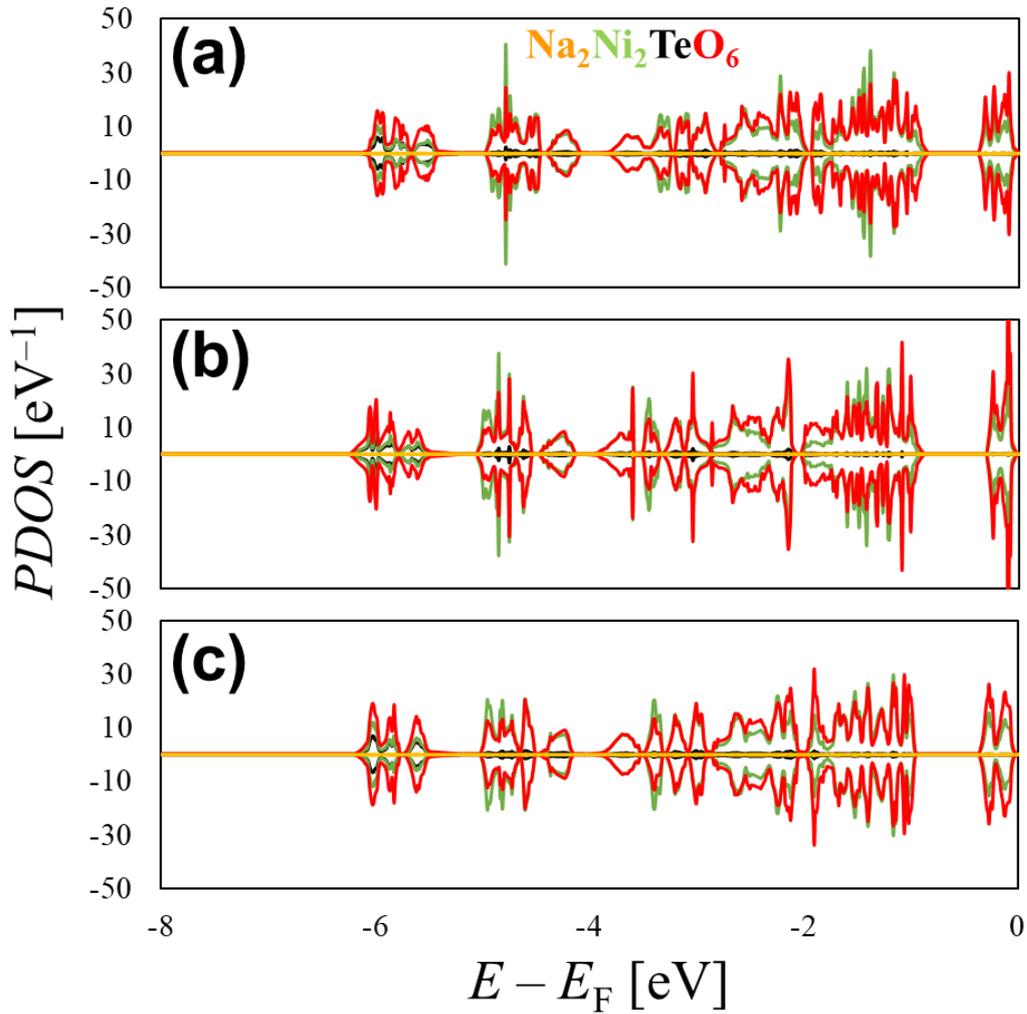

**Figure S2. Projected density of states (PDOS) plots of Na$_2$Ni$_2$TeO$_6$.** Orange, green, black, and red lines represent the PDOS of Na, Ni, Te, and O atoms, respectively. $E_F$ is the Fermi energy. PDOS plots of **(a)** Na$_2$Ni$_2$TeO$_6$ (for the structure shown in **Figures 2a** and **2b**), **(b)** Na$_2$Ni$_2$TeO$_6$ (for the structure shown in **Figures 2c** and **2d**) and **(c)** Na$_2$Ni$_2$TeO$_6$ (for the structure shown in **Figures 2e** and **2f**).

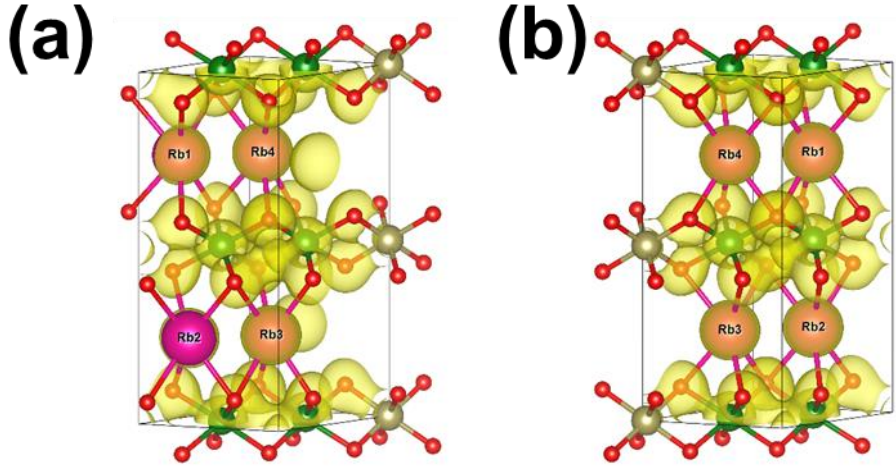

**Figure S3. Electron density distribution maps (yellow) of Rb$_2$Ni$_2$TeO$_6$.** Electron density distribution maps of **(a)** Rb$_2$Ni$_2$TeO$_6$ (for the structure shown in **Figures 3a** and **3b**) and **(b)** Rb$_2$Ni$_2$TeO$_6$ (for the structure shown in **Figures 3c** and **3d**). The threshold is 0.05 e$^-$/Bohr$^3$.

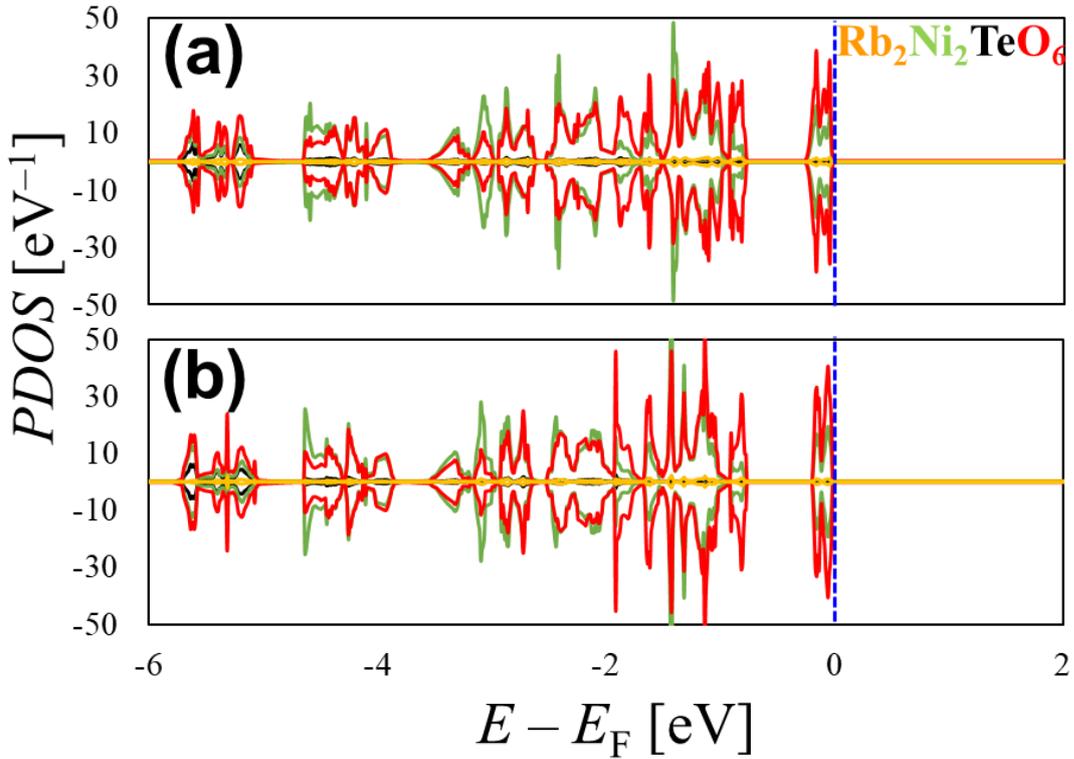

**Figure S4. Projected density of states (PDOS) plots of $Rb_2Ni_2TeO_6$.** Orange, green, black, and red lines represent the PDOS of Rb, Ni, Te, and O atoms, respectively. $E_F$ is the Fermi energy, the position of which is shown in blue broken lines. PDOS plots of **(a)** $Rb_2Ni_2TeO_6$ (for the structure shown in **Figures 3a** and **3b**) and **(b)** $Rb_2Ni_2TeO_6$ (for the structure shown in **Figures 3c** and **3d**).

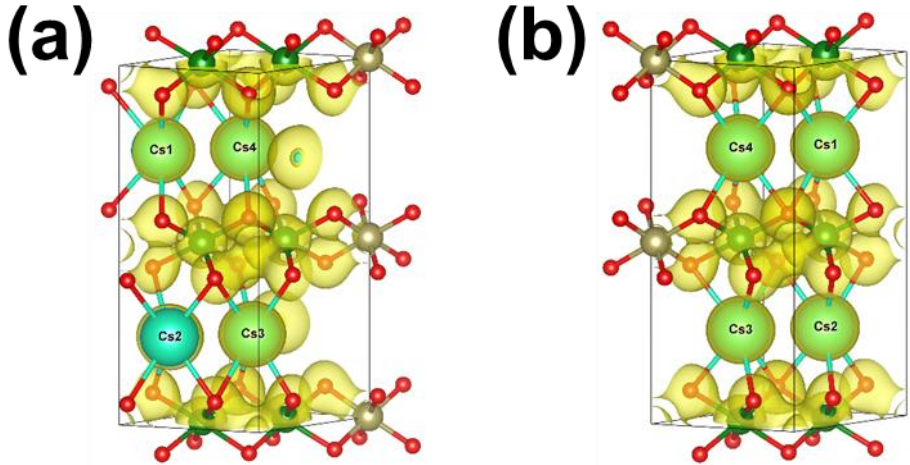

**Figure S5. Electron density distribution maps (yellow) of $Cs_2Ni_2TeO_6$.** Electron density distribution maps of **(a)** $Cs_2Ni_2TeO_6$ (for the structure shown in **Figures 4a** and **4b**) and **(b)** $Cs_2Ni_2TeO_6$ (for the structure shown in **Figures 4c** and **4d**). The threshold is 0.05 e$^-$/Bohr$^3$.

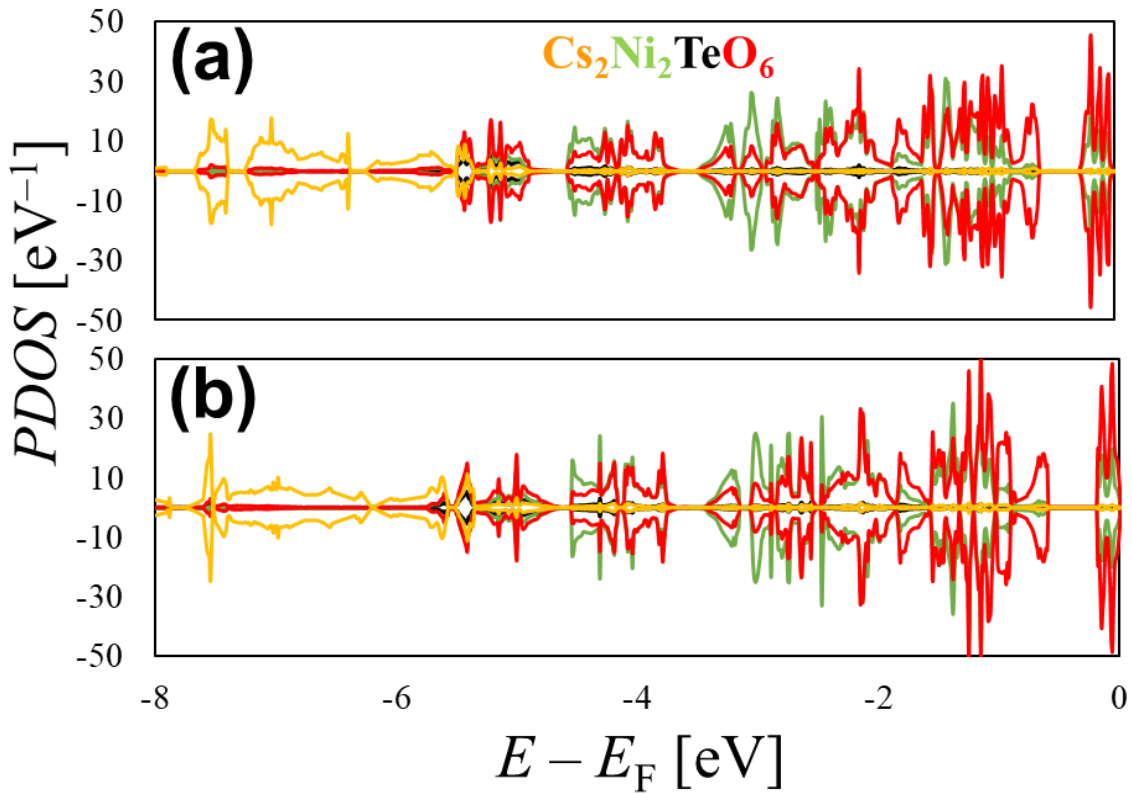

**Figure S6. Projected density of states (PDOS) plots of $Cs_2Ni_2TeO_6$.** Orange, green, black, and red lines represent the PDOS of Cs, Ni, Te, and O atoms, respectively. $E_F$ is the Fermi energy. PDOS plots of **(a)** $Cs_2Ni_2TeO_6$ (for the structure shown in **Figures 4a** and **4b**) and **(b)** $Cs_2Ni_2TeO_6$ (for the structure shown in **Figures 4c** and **4d**).

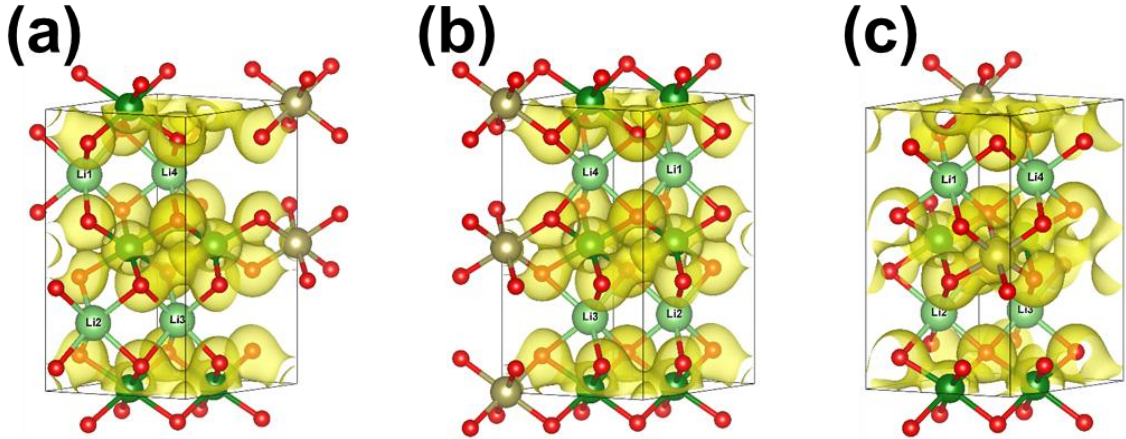

**Figure S7. Electron density distribution maps (yellow) of $Li_2Ni_2TeO_6$.** Electron density distribution maps of **(a)** $Li_2Ni_2TeO_6$ (for the structure shown in **Figures 5a** and **5b**), **(b)** $Li_2Ni_2TeO_6$ (for the structure shown in **Figures 5c** and **5d**) and **(c)** $Li_2Ni_2TeO_6$ (for the structure shown in **Figures 5e** and **5f**). The threshold is 0.05 e$^-$/Bohr$^3$.

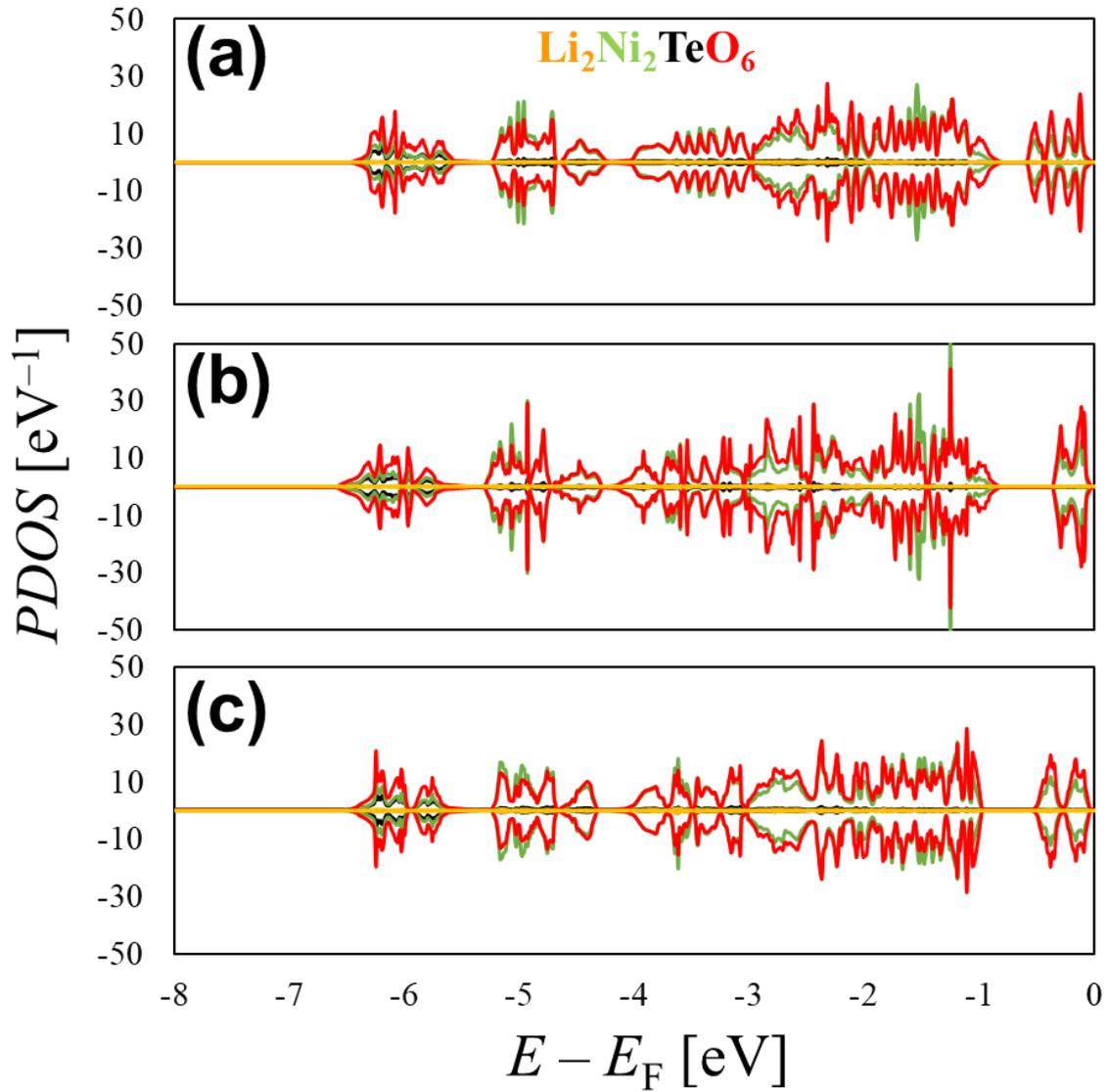

**Figure S8. Projected density of states (PDOS) plots of Li$_2$Ni$_2$TeO$_6$.** Orange, green, black, and red lines represent the PDOS of Li, Ni, Te, and O atoms, respectively. $E_F$ is the Fermi energy. PDOS plots of **(a)** Li$_2$Ni$_2$TeO$_6$ (for the structure shown in **Figures 5a** and **5b**), **(b)** Li$_2$Ni$_2$TeO$_6$ (for the structure shown in **Figures 5c** and **5d**) and **(c)** Li$_2$Ni$_2$TeO$_6$ (for the structure shown in **Figures 5e** and **5f**).

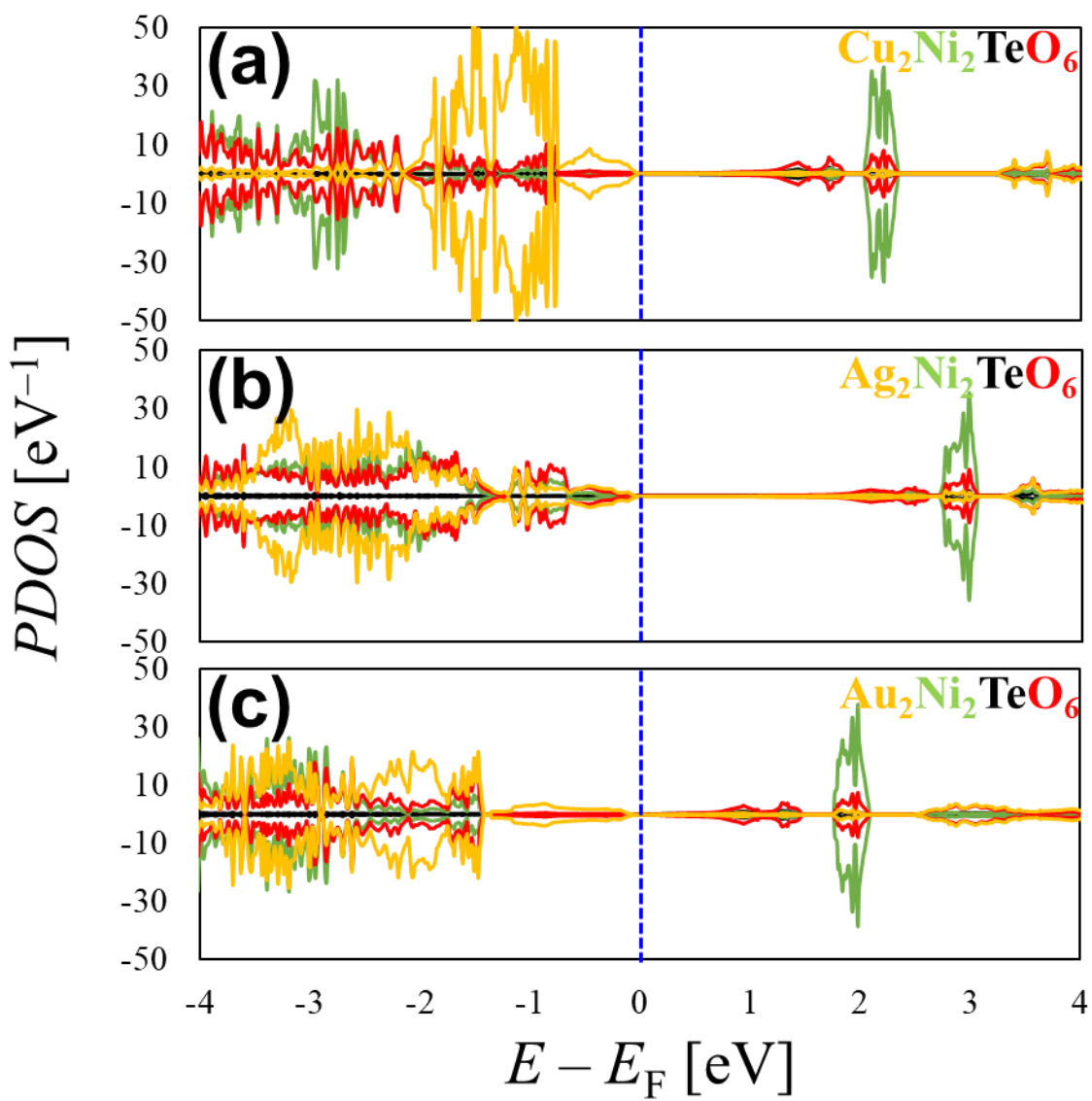

**Figure S9. Projected density of states (PDOS) plots of most stable structures of $A_2Ni_2TeO_6$ ($A$=Cu, Ag, Au).** Orange, green, black, and red lines represent the PDOS of $A$(=Cu, Ag, Au), Ni, Te, and O atoms, respectively. $E_F$ is the Fermi energy, the position of which is shown in blue broken lines. PDOS plots of **(a)** $Cu_2Ni_2TeO_6$ (for the structure shown in **Figures 7a** and **7b**), **(b)** $Ag_2Ni_2TeO_6$ (for the structure shown in **Figures 7c** and **7d**) and **(c)** $Au_2Ni_2TeO_6$ (for the structure shown in **Figures 7e** and **7f**).

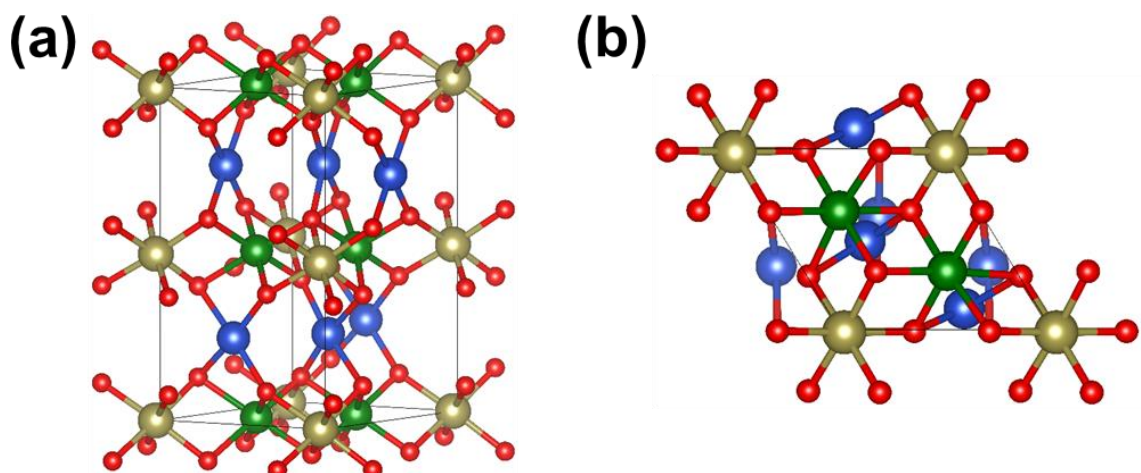

**Figure S10. Crystal structural frameworks of $Cu_2Ni_2TeO_6$ with Cu atoms (in blue) coordinated to four oxygen atoms.** Metastable crystal structural framework of $Cu_2Ni_2TeO_6$ with Cu in a planar coordination with oxygen atoms when viewed along **(a)** [110] zone axis and **(b)** [001] zone axis. Ni atoms (dark green) are arranged in a honeycomb configuration around Te atoms (in ochre) via the oxygen atoms (in red) to form $NiO_6$ and $TeO_6$ octahedra. Note that the crystal structural framework is slightly deviated from the [110] zone axis, in order to explicitly visualise all the atom coordination within the unit cell.

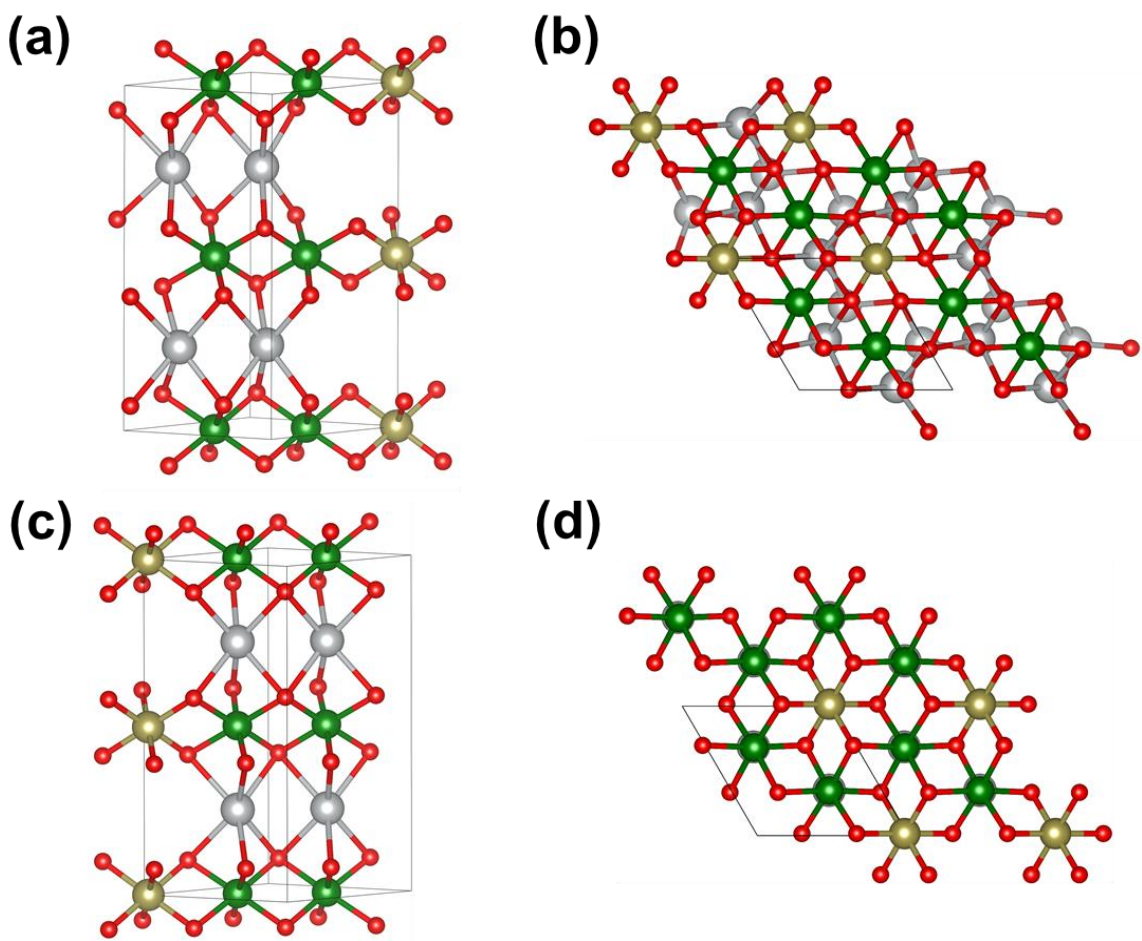

**Figure S11. Crystal structural frameworks of $Ag_2Ni_2TeO_6$ with Ag atoms (in grey) coordinated to six oxygen atoms.** Metastable crystal structural frameworks of $Ag_2Ni_2TeO_6$ with Ag in prismatic coordination with oxygen atoms when viewed along [110] zone axis (**(a)** and **(c)**) and [001] zone axis (**(b)** and **(d)**). Ni atoms (dark green) are arranged in a honeycomb configuration around Te atoms (in ochre) via the oxygen atoms (in red) to form $NiO_6$ and $TeO_6$ octahedra. Note that the crystal structural framework is slightly deviated from the [110] zone axis, in order to explicitly visualise all the atom coordination within the unit cell.

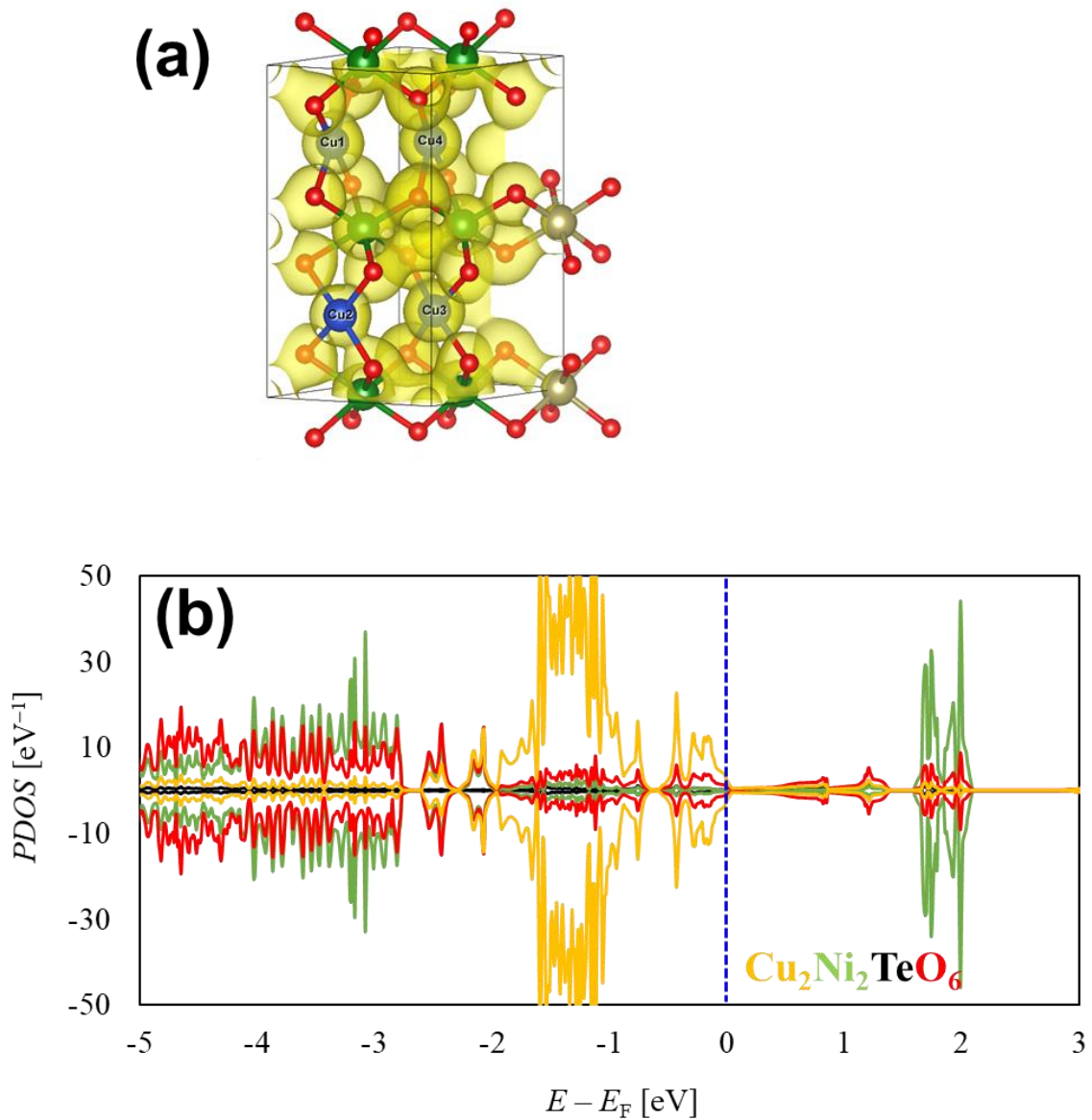

**Figure S12. Electron density distribution map and PDOS plots of a metastable structural framework of $Cu_2Ni_2TeO_6$ with Cu atoms coordinated to four oxygen atoms (in a square planar coordination).** (a) Electron density distribution (yellow) map of $Cu_2Ni_2TeO_6$ (for the structure shown in **Figure S10**). The threshold is 0.05 $e^-$/Bohr$^3$. (b) PDOS plots. Orange, green, black, and red lines represent the PDOS of Cu, Ni, Te, and O atoms, respectively. $E_F$ is the Fermi energy, the position of which is shown in blue broken lines.

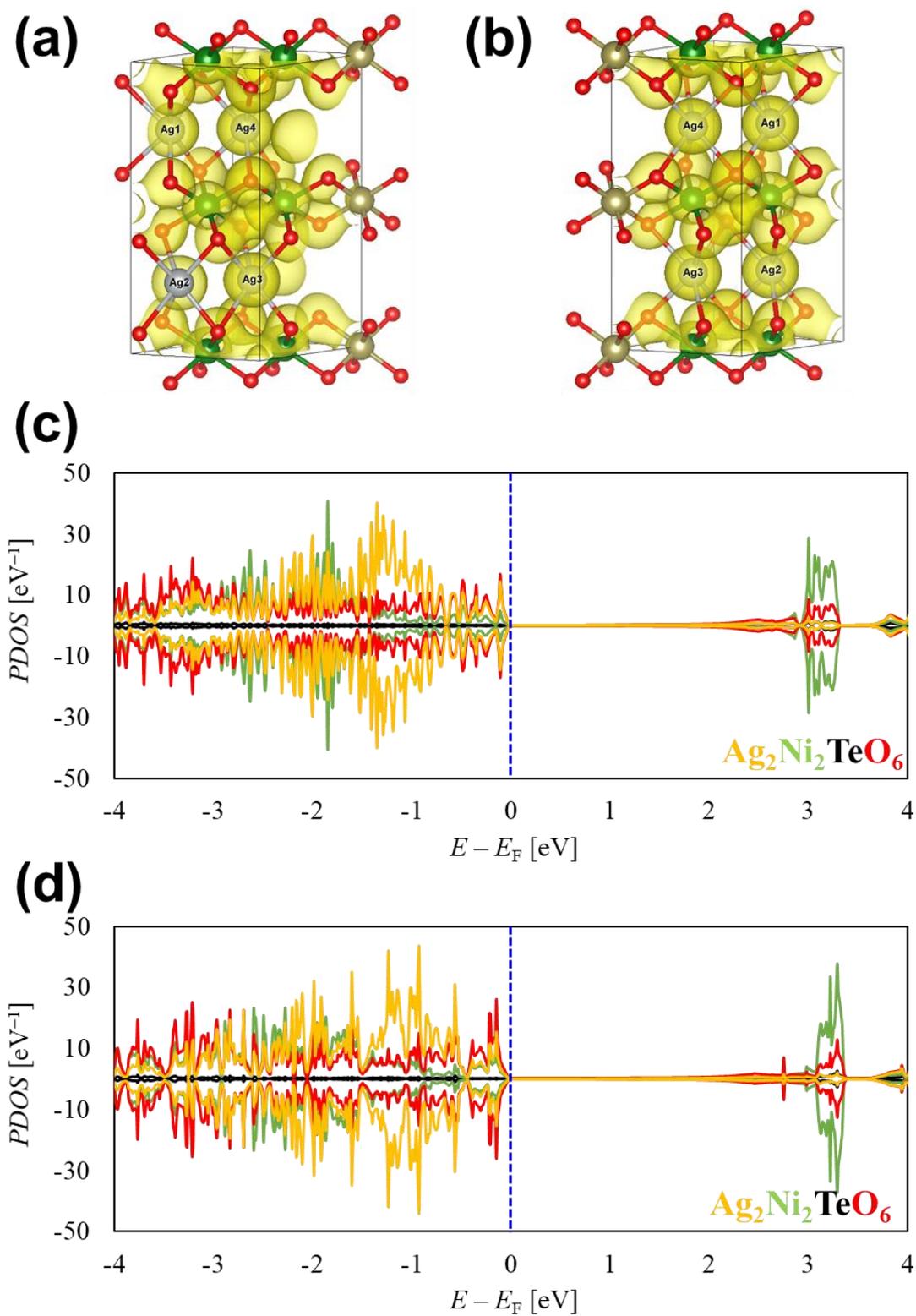

**Figure S13. Electron density distribution map and PDOS plots of metastable structural frameworks of $Ag_2Ni_2TeO_6$ with Ag atoms coordinated to six oxygen atoms in prismatic coordination.** (a) Electron density distribution (yellow) map of $Ag_2Ni_2TeO_6$ (for the structure shown in **Figures S11a** and **S11b**) and **(b)** $Ag_2Ni_2TeO_6$ (for the structure shown in **Figures S11c** and **S11d**). The threshold is 0.05 e$^-$/Bohr$^3$. **(c)** PDOS plots for the structure shown in **Figures S11a** and **S11b**. **(d)** PDOS plots for the structure shown in **Figures S11c** and **S11d**. Orange, green, black, and red lines represent the PDOS of Ag, Ni, Te, and O atoms, respectively. $E_F$ is the Fermi energy, the position of which is shown in blue broken lines.

## Data availability statement
The datasets (raw/processed data) required to reproduce the results and computer codes are available upon request from the authors.

## CRediT authorship contribution statement
**Kohei Tada:** Methodology, Investigation, Formal analysis, Data curation, Validation, Software, Resources, Funding acquisition, Writing – original draft. **Godwill Mbiti Kanyolo:** Methodology, Investigation, Validation, Writing - review & editing. **Titus Masese:** Methodology, Investigation, Validation, Visualization, Supervision, Funding acquisition,Writing - review & editing.

## Declaration of Competing Interests
The authors declare that they have no known competing financial interests or personal relationships that could have unethically impacted the rigour and scientific methods employed in this work.

## Acknowledgements
Computations described in this work were partially carried out using the computer facilities at the Research Institute for Information Technology, Kyushu University. This work was supported by the TEPCO Memorial Foundation. In addition, this work was also conducted in part under the auspices of the Japan Society for the Promotion of Science (JSPS KAKENHI Grant Numbers 21K14730 and 20K15177) and the National Institute of Advanced Industrial Science and Technology (AIST). T. Masese. and G. M. Kanyolo are grateful for the unwavering support from their family members (T. Masese.: Ishii Family, Sakaguchi Family and Masese Family; G. M. Kanyolo: Ngumbi Family). The authors also acknowledge the rigorous proofreading work on the manuscript done by Edfluent.

## Supplementary material

Supplementary material is available at the link given below:

https://ars.els-cdn.com/content/image/1-s2.0-S0927025622001124-mmc1.pdf